\documentclass[12pt,preprint]{aastex}

\newcommand{\bm}[1]{\mbox{\boldmath{$#1$}}}

\newcommand{\at}{{\rm Athena}}
\newcommand{\del}{{\bf \nabla}}

\newcommand{\etot}{E_{\rm tot}}
\newcommand{\md}{\dot{M}}
\newcommand{\kd}{\dot{K}}
\newcommand{\td}{\dot{T}}
\newcommand{\gd}{\dot{G}}
\newcommand{\qk}{\dot{Q}_{\rm k}}
\newcommand{\qm}{\dot{Q}_{\rm m}}
\newcommand{\ein}{E_{\rm in}}
\newcommand{\alf}{{\rm Alfv\acute{e}n}}
\newcommand{\pr}{P_m}

\newcommand{\maxwell}{\langle\langle-B_xB_y\rangle\rangle}
\newcommand{\reynolds}{\langle\langle \rho v_x\delta v_y\rangle\rangle}
\newcommand{\magnetic}{\langle\langle B^2/2 \rangle\rangle}
\newcommand{\magx}{\langle\langle B_x^2/2 \rangle\rangle}
\newcommand{\magy}{\langle\langle B_y^2/2 \rangle\rangle}
\newcommand{\magz}{\langle\langle B_z^2/2 \rangle\rangle}

\newcommand{\pertkinetic}{\langle\langle \rho \delta v^2/2\rangle\rangle}
\newcommand{\kinx}{\langle\langle \rho v_x^2/2\rangle\rangle}
\newcommand{\pertkiny}{\langle\langle \rho \delta v_y^2/2\rangle\rangle}
\newcommand{\kinz}{\langle\langle \rho v_z^2/2\rangle\rangle}

\newcommand{\bstar}{\widetilde{{\bm B}}^{\bm \ast}({\bm k})}
\newcommand{\bfour}{\widetilde{{\bm B}}({\bm k})}
\newcommand{\vstar}{\widetilde{\sqrt{\rho}{\bm v}^{\bm \ast}}({\bm k})}
\newcommand{\vfour}{\widetilde{\sqrt{\rho}{\bm v}}({\bm k})}
\newcommand{\vpfour}{\widetilde{\sqrt{\rho}{\bm \delta}{\bm v}}({\bm k})}
\newcommand{\imi}{{\it i}}
\newcommand{\vsh}{V_{{\rm sh}}}
\newcommand{\vt}{{\bm v_{\rm t}}}
\newcommand{\dv}{{\rm d^3}{\bm x}}
\newcommand{\sr}{\sqrt{\rho}}
\newcommand{\fterm}{{\rm e}^{-\imi {\bm k} \cdot {\bm x}} \dv}
\newcommand{\etaeff}{\eta_{\rm eff}}
\newcommand{\nueff}{\nu_{\rm eff}}
\newcommand{\reeff}{Re_{\rm eff}}
\newcommand{\rmeff}{Rm_{\rm eff}}
\newcommand{\pmeff}{P_{m,{\rm eff}}}

\newcommand{\tbb}{T_{bb}}
\newcommand{\tdivv}{T_{{\rm div}v}}
\newcommand{\tbv}{T_{bv}}
\newcommand{\tvv}{T_{vv}}
\newcommand{\tcomp}{T_{\rm comp}}
\newcommand{\tvb}{T_{vb}}
\newcommand{\tpress}{T_{\rm press}}
\newcommand{\tcor}{T_{\rm cor}}
\newcommand{\tphi}{T_\phi}
\newcommand{\dmag}{D_{\rm mag}}
\newcommand{\dkin}{D_{\rm kin}}
\newcommand{\tnu}{T_\nu}
\newcommand{\teta}{T_\eta}

\newcommand{\zonesix}{{\rm SZ16}}
\newcommand{\zthree}{{\rm SZ32}}
\newcommand{\zsix}{{\rm SZ64}}
\newcommand{\zone}{{\rm SZ128}}
\newcommand{\nonesix}{{\rm NZ16}}
\newcommand{\nthree}{{\rm NZ32}}
\newcommand{\nsix}{{\rm NZ64}}
\newcommand{\none}{{\rm NZ128}}

\newcommand{\zdone}{{\rm SZD128}}

\newcommand{\ndone}{{\rm NZD128}}

\newcommand{\narone}{{\rm NZAR1}}
\newcommand{\nartwo}{{\rm NZAR2}}
\newcommand{\narthree}{{\rm NZAR3}}
\newcommand{\narfour}{{\rm NZAR4}}
\newcommand{\narfive}{{\rm NZAR5}}
\newcommand{\narsix}{{\rm NZAR6}}

\begin{document}

\title{Simulations of Magnetorotational Turbulence with a Higher-Order Godunov Scheme}

\author{Jacob B. Simon, John F. Hawley, Kris Beckwith}
\affil{Department of Astronomy \\ University of Virginia \\ P.O. Box 400325 \\
Charlottesville, VA 22904-4325}

\begin{abstract}
We apply a new, second-order Godunov code, Athena, to studies of the
magnetorotational instability (MRI) using unstratified shearing box
simulations with a uniform net vertical field and a sinusoidally
varying zero net vertical field.  The Athena results agree well with
similar studies that used different numerical algorithms, including
the observation that the turbulent energy decreases with increasing
resolution in the zero net field model.  We conduct analyses to study the
flow of energy from differential rotation to turbulent fluctuations
to thermalization.  A study of the time-correlation between the rates
of change of different volume-averaged energy components shows that
energy injected into turbulent fluctuations dissipates on a timescale
of $\Omega^{-1}$, where $\Omega$ is the orbital frequency of the local
domain.  Magnetic dissipation dominates over kinetic dissipation, although
not by as great a factor as the ratio of magnetic to kinetic energy.
We Fourier-transform the magnetic and kinetic energy evolution equations
and, using the assumption that the time-averaged energies are constant,
determine the level of numerical dissipation as a function of length
scale and resolution.  By modeling numerical dissipation as if it were
physical in origin, we characterize numerical resistivity and viscosity
in terms of effective Reynolds and Prandtl numbers.  The resulting
effective magnetic Prandtl number is $\sim 2$, independent of resolution
or initial field geometry.  MRI simulations with effective Reynolds and
Prandtl numbers determined by numerical dissipation are not equivalent to
those where these numbers are set by physical resistivity and viscosity.
These results serve, then, as a baseline for future shearing box studies
where dissipation is controlled by the inclusion of explicit viscosity
and resistivity.
\end{abstract}


\section{Introduction}
\label{introduction}

The process of accretion powers a wide range of astrophysical systems,
from protostars to quasars.  In accretion disks, gravitational energy is
converted into other forms including bulk outflows, heat, and radiation.
In the traditional time-stationary thin disk model of \cite{shak73}, the
$r,\phi$ component of the stress, $\tau_{r\phi}$, is proportional to the
local pressure, $\tau_{r\phi} = \alpha P$.  The $\alpha$ model assumes
that the accretion energy is deposited as heat locally and radiated
rapidly, providing a relation between disk emissivity and accretion rate.
While the $\alpha$ model has proven valuable in interpreting many aspects
of accretion systems, advancing beyond it will require a more detailed
understanding of the stress that produces angular momentum
transport as well as the physical processes involved in the subsequent
thermalization and radiation of the orbital energy released by those
stresses.

It is now understood that magnetohydrodynamic (MHD) turbulence generated
by the magnetorotational instability (MRI) \cite[]{balb91,balb98} produces
significant Maxwell stresses, $-B_rB_\phi/4\pi$, and Reynolds stresses,
$\rho \delta v_r \delta v_\phi$, that account for transport within
accretion disks.  The absence of an analytic theory for MHD turbulence,
however, means that direct numerical simulations play an essential role
in investigating accretion physics.  In this regard, local simulations,
which reduce the problem to the simplest form that can sustain MRI-driven
turbulence, have proven very useful.  The ``shearing box'' model is
a representation of a small patch of the disk constructed by boosting to
a local co-rotating Cartesian frame that ignores geometric curvature but
retains all rotational forces.  MRI shearing box simulations were
introduced by \cite{haw95} and have been extensively used since then
both without \cite[e.g.,][]{haw96,balb98} and with vertical stratification
\cite[e.g.,][]{bran95,stone96,hir06}.

Shearing box simulations can investigate several key questions including
the functional dependence of the stress on disk properties
and the turbulent energy flow that leads to dissipation as heat.
These simulations have made it increasingly clear, for example, that
the basic $\alpha$ stress parameterization is not only too simplistic,
it is actually misleading.  Shearing boxes have provided ample evidence
that stress is {\it not} determined by pressure, at least in the usual
manner of the $\alpha$ disk \cite[][]{haw95, sano04}. Early studies showed
instead that stress is (in some cases) proportional to the {\it magnetic} pressure,
but the magnetic energy is not itself directly determined by the gas and
radiation pressure.  \cite{black08} recently reviewed a large number of
shearing box results and found that this result holds across the full
ensemble of simulations with only small differences in the constant of
proportionality from one run to another.  The implications of these results are
significant.  For example, recent local simulations using stratified
shearing boxes and radiation transport \cite[][]{bla07, kro07}
have found no evidence of the thermal instability long believed to be
present in radiation-pressure supported $\alpha$ disks.

If stress is proportional to magnetic rather than total pressure,
what determines the magnetic pressure in a disk?  Apart from the
expectation that the field will remain subthermal, this remains uncertain.
The simplest shearing box simulations using ideal MHD have a limited
range of significant parameters; this is both a strength and a weakness
of that model.  The magnetic energy in the saturated state could depend
upon such factors as box size, the amplitude and geometry of the imposed initial
magnetic field, and the ratio of the gas pressure to magnetic pressure
(the plasma $\beta$ value).  \cite{haw95} and \cite{haw96} studied the
effect of initial magnetic field topology on the resulting stress  and
found that although the MRI leads to turbulence regardless of the initial
field, simulations that had an imposed net vertical field produce higher
turbulence levels than an imposed toroidal field or a simulation that
began with zero net magnetic flux within the domain.  \cite{haw95} found
that the total magnetic energy and the resulting stress in the saturated
turbulent state was a function of the initial plasma $\beta$ with a
uniform vertical field, namely that larger $\beta$ (i.e., weaker fields)
leads to smaller saturation levels.  Other initial field configurations
do not yield so direct a correlation between background field strength
and saturation.  Many simulations have failed to find any noticeable
correlation between mean turbulent magnetic energy and the gas pressure.
A comprehensive parameter study by \cite{sano04} observed at best only
a very weak gas pressure dependence.

Since the mean magnetic energy at saturation is presumably a balance
between continued driving by the MRI and loss due to magnetic dissipation
and reconnection, there has been interest in going beyond ideal MHD to
include explicit physical dissipation in the form of kinematic viscosity,
$\nu$, and Ohmic resistivity, $\eta$.  Both of these properties have been shown to be
important in determining the mean energies and stresses in MRI turbulence.
Simulations by \cite{haw96}, \cite{sano98}, \cite{flem00}, \cite{sano01},
\cite{zieg01}, and \cite{sano02b} have investigated the impact of a
nonzero $\eta$.  The main result of these studies is that increasing
the resistivity leads to a decrease in turbulence, independent of the
initial field configuration.  In zero net field models, the effect of
resistivity on the turbulence is larger than one might expect from the
linear MRI relation \cite[]{flem00}.  On the other hand, \cite{haw96}
found that increasing the viscosity increased the magnetic energy in the
saturated state.  Recent work has clarified the situation by demonstrating
a dependence of the saturation level on both $\eta$ and $\nu$ in terms
of the magnetic Prandtl number, $\pr = \nu/\eta$.  In particular, the
level of angular momentum transport increases with increasing $\pr$ for
simulations initiated with a uniform as well as vanishing mean magnetic
field in the vertical direction \cite[]{from07b,lesur07}.

Determining the stress levels in MRI turbulence is only one aspect of
the problem; another is exploring how that turbulence is dissipated
into heat.  This question has direct relevance to phenomenological disk
models as well as observations.  The $\alpha$ model
assumes that the accretion energy is deposited as heat locally and
rapidly, and \cite{balb99} showed that this property should hold for 
the energetics of MHD turbulence as well.
In the simulations, we can determine the rate at which turbulent energy
is thermalized and the path that energy takes as it moves from the free
energy of the shear flow to turbulence and then to heat.  Such issues
were briefly touched on by \cite{bran95} who found that the turbulent
magnetic energy was $\sim 6$ times greater than the perturbed kinetic
energy, but dissipational heating resulted from roughly equal contributions of
magnetic and kinetic energy dissipation.  This result led them to suggest
that there was a net transfer of magnetic energy to turbulent kinetic energy.  
\cite{sano01} studied
energy flow in the context of MRI channel modes, which are strong radial
streaming motions that result from the linear growth of the vertical
field MRI \cite[]{haw92,balb98}.  Their work included Ohmic resistivity
(but not viscosity) and showed that resistive heating dominated the
thermalization of energy stored in these channel modes. Dissipational
heating also plays an important role in radiative effects and determining
disk structure, both of which may be observable properties of disks \cite[e.g.,][]{beck08}.

In any study that depends on simulations, there remain
factors which cannot be overlooked:  the effects due to numerics and
finite resolution.  The majority of the results to-date were obtained
with numerical codes based on the finite-difference ZEUS algorithm
\cite[]{stone92a,stone92b}, carried out at relatively low resolution.
ZEUS is effectively first-order in asymptotic convergence, and in
its most widely used form, evolves the internal rather than the
total energy equation.  There have been improvements in both the
available computational power, which makes higher resolutions and longer
evolution times possible, and in the algorithms for compressible MHD.
In this work, we will reexamine the properties of MHD turbulence in the
shearing box using a higher-order, Godunov scheme.

The new code, $\at$, \cite[see][]{stone08} represents an improvement
over ZEUS in several ways including true second-order convergence,
increased effective resolution \cite[see][]{stone05}, accurate shock capturing,
and conservation of total energy.  The energy-conserving properties
of $\at$ allow us to study energy flow and
dissipation within the shearing box in greater detail than allowed for
by the ZEUS algorithm.  The version of $\at$ we use in this paper does
not include explicit resistivity or viscosity and instead relies on
numerical dissipation to thermalize the turbulent energy.  Nevertheless,
this work will serve as a starting point for planned studies of nonideal
effects, including the influence of $\pr$ on the turbulence
\cite[][]{from07b, lesur07}.  As an important
part of establishing a baseline of simulations, we will characterize
the numerical resistivity and viscosity of $\at$ for the shearing box
problem.  To do so, we will follow the recent work of \cite{from07a} who
studied the numerical effects of ZEUS on the saturated state of MRI shearing
box simulations that begin with zero net field.  They found that the
amplitude of the turbulence decreases with increasing resolution and
developed several useful diagnostics with which to quantify the effective
numerical resistivity and viscosity in the problem.

The structure of the paper is as follows.  In \S~\ref{method},
we describe the algorithm employed and our simulations.  In
\S~\ref{general_properties}, we reexamine some of the 
results from previous MRI studies and provide a comparison with these
studies.  In \S~\ref{energy_fluctuations}, we present the first of two
diagnostics used to study turbulent energy flow and dissipation.  The
second of these diagnostics is applied in \S~\ref{trans_funcs}.  Finally,
we discuss our results and summarize our conclusions in \S~\ref{conclusions}.

\section{Numerical Simulations}
\label{method}

The code used for all of our simulations is $\at$, a second-order accurate
Godunov scheme for solving the equations of ideal MHD in conservative
form.  The equations are solved using the dimensionally unsplit corner
transport upwind (CTU) method of \cite{col90} coupled with the third-order in space
piecewise parabolic method (PPM) of \cite{col84} and a constrained
transport (CT) algorithm for preserving the $\del \cdot {\bm B}$~=~0
constraint.  Details of the algorithm are described in \cite{gard05a},
\cite{gard08}, and \cite{stone08}.  The $\at$ code has been extensively
tested against various hydrodynamic and MHD tests \cite[]{stone08}.

We employ the shearing box formalism, in which our computational domain
is corotating with the fluid flow at some radius in the disk.  The domain
size is small compared to this radius, allowing us to expand the equations
of motion in Cartesian form, as described in detail by \cite{haw95}.

In the ideal MHD approximation, the evolution of the fluid in the
shearing box is described by:

\begin{equation}
\label{cont}
\frac{\partial \rho}{\partial t} + \nabla \cdot (\rho {\bm v}) = 0,
\end{equation}
\begin{equation}
\label{momentum}
\frac{\partial \rho {\bm v}}{\partial t} + \nabla \cdot (\rho {\bm v}{\bm v} - {\bm B}{\bm B}) + 
\nabla (P + \frac{1}{2} B^2) = 2 q \rho \Omega^2 {\bm x} - 2 {\bm \Omega} \times \rho {\bm v},
\end{equation}
\begin{equation}
\label{induction}
\frac{\partial {\bm B}}{\partial t} + \nabla \cdot ({\bm v}{\bm B} - {\bm B}{\bm v}) = 0,
\end{equation}
\begin{equation}
\label{energy_eqn}
\frac{\partial E}{\partial t} + \nabla \cdot [(E + P + \frac{1}{2} B^2) {\bm v} 
- {\bm B}({\bm B} \cdot {\bm v})] = 2 q \Omega^2 \rho {\bm v} \cdot {\bm x},
\end{equation}

\noindent 
where $\rho$ is the mass density, $\rho {\bm v}$ is the
momentum density, ${\bm B}$ is the magnetic field, $P$ is the gas
pressure, $E$ is the total energy density, and $q$ is the shear
parameter, defined as $q = -d$ln$\Omega/d$ln$R$. $\Omega$ is the angular velocity of the
center of the shearing box.  Note that our system
of units has the magnetic permeability $\mu = 1$.  We use $q = 3/2$,
appropriate for a Keplerian disk. The first source
term on the right-hand side of equation~(\ref{momentum}) and the term on
the right-hand side of equation~(\ref{energy_eqn}) correspond to
tidal forces (gravity and centrifugal) in the corotating frame.  
The second source term in
equation~(\ref{momentum}) is the Coriolis force. The total energy density
is the sum of the thermal, kinetic, and magnetic energy
densities

\begin{equation}
\label{energy}
E = \epsilon + \frac{1}{2} \rho v^2 + \frac{1}{2} B^2
\end{equation}

\noindent 
where $\epsilon$ is thermal energy density.  The equation of
state is that of an ideal gas, $\epsilon = P / (\gamma-1)$,
where the adiabatic index is $\gamma = 5/3$ in all simulations. The
terms on the right hand sides of equations~(\ref{momentum}) and
(\ref{energy_eqn}) are added to the MHD equations in a directionally
unsplit manner, consistent with the CTU algorithm.  Note that we have
neglected vertical stratification.

An important component of shearing box simulations is the shearing
periodic boundary conditions at the $x$ boundaries, which are implemented
as described in \cite{haw95} with a few modifications for $\at$.  First,
as in \cite{haw95}, the $y$ momentum is adjusted to account for the
shear across the $x$ boundaries as fluid moves out one boundary and
enters at the other.  Since $\at$ evolves the total energy, however,
this energy must also be adjusted to account for the difference in $y$
momentum across the boundaries.  Second, following the description in
\cite{haw95}, quantities are linearly reconstructed in the
ghost zones from appropriate zones in the physical domain that have
been shifted along $y$ to account for the shear across the boundary.
However, we have found that the precise conservation of a quantity
depends on how this reconstruction is performed; the fluxes of a
particular conserved quantity must be reconstructed to conserve
the quantity to roundoff level.
For example, consider the conservation of magnetic flux through the computational domain.
For the magnetic flux through the box to be
conserved to machine precision, the line integral of the electromotive
force (EMF), ${\cal E} = -{\bm v} \times {\bm B}$, along the boundaries
must remain zero.  The $y$ and $z$ boundary conditions are periodic,
and therefore, the line integrated EMFs along these boundaries cancel.
This is not the case with the shearing periodic boundaries, however.  Consider 
the net $B_z$ flux through the grid, which will be conserved
if ${\cal E}_y = v_z B_x - v_x B_z$ is zero when integrated along both
$x$ boundaries.  Computing the EMF using ghost zone variables $v_z,
B_x, v_x,$ and $B_z$ after reconstruction introduces a truncation error,
and the $B_z$ flux is not conserved.  This is avoided if we instead
perform the shearing-periodic reconstruction step on ${\cal E}_y$ itself.
A similar argument applies to mass conservation; one needs to reconstruct
the density flux in the shearing boundaries instead of the density itself.\footnote{In principle, 
the same argument applies to momentum and energy conservation, but
these equations are not conserved to machine precision due to the
existence of source terms.}

In the code, we only perform this EMF/flux reconstruction for ${\cal E}_y$.
We have found that conservation of the $B_z$ flux is essential, owing
to the strong effect on the turbulence due to a net vertical field.
The perfect conservation of $B_y$ is not as important, and as ensuring
its precise conservation involves a more complex procedure, we allow
the $B_y$ flux to be conserved only to the truncation level.  Similarly,
the precise conservation of mass has minimal impact on the behavior
of the turbulence, and we allow the mass to be conserved to truncation
level.  We would like to note, however, that because of this, mass is lost
during the MRI evolution in our simulations.  To quantify the level of
mass loss, the total percentage of mass lost over 100 orbits of evolution 
is $\sim$ 2\% for our highest resolution simulations (see below for a description of our
simulations) and $\sim$ 10\% for our lowest resolution simulations; 
we observe convergence of mass conservation with resolution.

Although $\at$ conserves total energy, the shearing
boundaries do work on the fluid and represent a significant energy
source.
As was shown in \cite{haw95}, one can integrate the total energy plus
gravitational potential energy, $E + \rho \Phi$, where $\Phi = q \Omega^2
(\frac{L_x^2}{12}-x^2)$, over the domain to obtain

\begin{equation}
\label{ein}
\frac{\partial \langle E+\rho \Phi\rangle}{\partial t} = \frac{q \Omega}{L_y L_z} \int_X (\rho v_x \delta v_y - B_x B_y) dy dz, 
\end{equation}

\noindent
where $L_x$, $L_y$, and $L_z$ are the domain sizes in the $x$, $y$, and $z$ directions respectively (see below),

\begin{equation}
\label{pert_vy}
\delta v_y \equiv v_y+q\Omega x,
\end{equation}

\noindent 
and the integral is calculated over one of the $x$ boundaries.
In our simulations, equation~(\ref{ein}) is satisfied to truncation 
level with the error coming from the tidal potential source term
in equation~(\ref{energy_eqn}).  It is possible to rewrite this source
term to guarantee that equation~(\ref{ein}) is satisfied to roundoff
level \cite[see][]{gard05b}, but we have found that this makes very
little difference to how the total energy evolves.

\cite{gard05b} point out that the source terms in the momentum equation
cannot be written in a purely conservative form and that the $x$ and $y$
momenta are tightly coupled through these terms.  In the hydrodynamic
limit the source terms account for epicyclic oscillations, and if the
epicyclic kinetic energy (see their equation~8) is not conserved to
machine precision, coupling between long wavelength modes and epicyclic
oscillation modes can result from truncation error.  Over time this
coupling can artificially increase the kinetic energy.  To ensure the
conservation of epicyclic energy, \cite{gard05b} evolved the angular
momentum fluctuations directly rather than the $y$ momentum, casting
the equations into a form consistent with uniform epicyclic motion.
They then employed a Crank-Nicholson scheme to evolve the source terms
that govern the evolution of the mometum fluctuations.  In MHD, however,
oscillatory epicyclic motion is replaced by unstable, growing MRI modes.
Epicyclic kinetic energy is not conserved and these special techniques
are not required.  Therefore, we use the standard Athena algorithm
\cite[e.g.,][]{stone08} to evolve the momentum equations.

As was done in the original shearing box simulations \cite[]{haw95} our
standard shearing box has a radial size $L_x = 1$, an azimuthal size $L_y
= 2\pi$, and a vertical size $L_z = 1$.  We initialize a velocity flow with
${\bm v} = -q\Omega x \hat{{\bm y}}$, with $q$~=~3/2, $\Omega$~=~0.001,
and $-L_x/2 \leq x \leq L_x/2$.  In an isothermal disk, the sound speed
is $c_{\rm s} \sim \Omega H$ where $H$ is the scale height.  With $L_z = H$,
we have $c_{\rm s}$~=~$L_z\Omega$, and we define the initial pressure as $P =
\rho \Omega^2 L_z^2$.  With $\rho$~=~1, we have $P$~=~$10^{-6}$.  In this
paper, we consider two initial magnetic field geometries that are commonly
used in shearing box studies.  Models labeled NZ (for Net Z-field) have
an initial uniform vertical magnetic field, $B_z$, and models labeled SZ
(for Sine Z-field) begin with a sinusoidal distribution of $B_z$ and
have zero net flux through the box.  Specifically, we initialize the NZ
runs with ${\bm B} = \sqrt{2P/\beta}\,\hat{{\bm z}}$, and the SZ runs
with ${\bm B} = \sqrt{2P/\beta}\,{\rm sin[}(2\pi/L_x) x{\rm]}\hat{{\bm
z}}$.  In both cases, we set $\beta =1600$.  This determines the ratio
of the vertical box size to the fastest growing linear MRI wavelength as
$L_z/\lambda_c \sim$~4, where $\lambda_c = 2\pi \sqrt{16/15}|v_{\rm A}|/\Omega$,
and $v_{\rm A}$ is the $\alf$ speed.  To seed the MRI, we introduce random
adiabatic perturbations to $P$ and $\rho$ with amplitude $\delta
P/P$~=~0.01.

For both of these initial field configurations, we have run a full range of
grid resolutions, from $N_x$~=~16, $N_y$~=~32, $N_z$~=~16 to the highest
resolution used in this study, $N_x$~=~128, $N_y$~=~256, $N_z = 128$,
proceeding by factors of two.  All of the simulations were run for a total
of 100~orbits.

In addition to the standard shearing box simulations, we have run some
additional experiments designed to further investigate magnetic and
kinetic energy dissipation.  First, we perform a set of simulations
in which we remove the velocity shear and the tidal and Coriolis force
terms, thus removing the energy source that maintains the turbulence.
The purpose of these simulations is to investigate energy flow and
dissipation in the absence of the shear, which is the driving force for
the turbulence.  We perform these simulations by restarting each of the
standard shearing box runs at a time when the shearing periodic boundaries
are strictly periodic. These ``periodic points" are given by $t_n =
n\,L_y/q\,\Omega\,L_x$, with $n = 0, 1, 2, ...,$ \cite[see][]{haw95}. We
choose the restart time to be 40~orbits.  We then evolve the system to
follow the decay of the kinetic and magnetic energies.

Finally, we run a set of low resolution simulations with varying aspect
ratio to examine the effect of secondary parasitic modes on the channel
solution (see \S~\ref{channel_solution}). These simulations have the
same initial conditions as the net flux simulation with $N_x = 32$,
$N_y = 64$, and $N_z = 32$ but with varying domain size in the $x$
and $y$ dimensions.  The grid cell size (e.g., $L_x/N_x$ in the $x$
direction) in each dimension is kept constant.  All simulations are
summarized in Table~\ref{tbl:runs}.

\clearpage
\begin{deluxetable}{l|cccc}
\tabletypesize{\scriptsize}
\tablewidth{0pc}
\tablecaption{MRI Simulations with $\at$ \label{tbl:runs}}
\tablehead{
\colhead{Label}&
\colhead{Initial Field Geometry}&
\colhead{Resolution ($N_x \times N_y \times N_z$)}&
\colhead{Domain ($L_x \times L_y \times L_z$)}&
\colhead{Description}    }
\startdata
$\nonesix$ & net flux & 16 $\times$ 32 $\times$ 16 & $1 \times 2\pi \times 1$ & -- \\
$\nthree$ & net flux & 32 $\times$ 64 $\times$ 32 & $1 \times 2\pi \times 1$ & -- \\
$\nsix$ & net flux & 64 $\times$ 128 $\times$ 64 & $1 \times 2\pi \times 1$ & -- \\
$\none$ & net flux & 128 $\times$ 256 $\times$ 128 & $1 \times 2\pi \times 1$ & fiducial run - net flux \\
$\ndone$ & net flux & 128 $\times$ 256 $\times$ 128 & $1 \times 2\pi \times 1$ & decaying turbulence \\
$\zonesix$ & zero net flux & 16 $\times$ 32 $\times$ 16 & $1 \times 2\pi \times 1$ & -- \\
$\zthree$ & zero net flux & 32 $\times$ 64 $\times$ 32 & $1 \times 2\pi \times 1$ & -- \\
$\zsix$ & zero net flux & 64 $\times$ 128 $\times$ 64 & $1 \times 2\pi \times 1$ & -- \\
$\zone$ & zero net flux & 128 $\times$ 256 $\times$ 128 & $1 \times 2\pi \times 1$ & fiducial run - zero net flux \\
$\zdone$ & zero net flux & 128 $\times$ 256 $\times$ 128 & $1 \times 2\pi \times 1$ & decaying turbulence \\
$\narone$ & net flux & 16 $\times$ 64 $\times$ 32 & $\frac{1}{2} \times 2\pi \times 1$ & varied aspect ratio \\
$\nartwo$ & net flux & 64 $\times$ 64 $\times$ 32 & $2 \times 2\pi \times 1$ & varied aspect ratio \\
$\narthree$ & net flux & 32 $\times$ 32 $\times$ 32 & $1 \times \pi \times 1$ & varied aspect ratio \\
$\narfour$ & net flux & 16 $\times$ 32 $\times$ 32 & $\frac{1}{2} \times \pi \times 1$ & varied aspect ratio \\
$\narfive$ & net flux & 64 $\times$ 32 $\times$ 32 & $2 \times \pi \times 1$ & varied aspect ratio \\
$\narsix$ & net flux & 128 $\times$ 32 $\times$ 32 & $4 \times \pi \times 1$ & varied aspect ratio \\
\enddata
\end{deluxetable}
\clearpage

\section{General Properties of MRI Turbulence}
\label{general_properties}

This work represents the first detailed study of the MRI with $\at$,
which has an algorithm significantly different from that used in ZEUS.
To begin, we will reexamine many of the shearing box models and the
results already documented in the literature.  Any significant differences
between $\at$ results and those previously published could indicate
where numerical effects (algorithm, resolution) have an influence.
Since $\at$ is an energy-conserving, shock-capturing algorithm it has at
least the potential to produce somewhat different results.  Conversely,
agreement between $\at$ and other codes would support the robustness of
the shearing box results to date.

In this section, we describe some of the general properties of MRI
turbulence as simulated with $\at$ and compare our results with those in
the literature.   These properties will also serve as a starting point for
further analysis presented in the following sections.  In what follows,
the highest resolution runs $\none$ and $\zone$ will serve as our fiducial
simulations for each initial field geometry.  We study resolution effects
for each field geometry using the lower resolution simulations.

\subsection{Characteristics of Saturation}
\label{sat_characteristics}

\begin{figure}
\plotfiddle{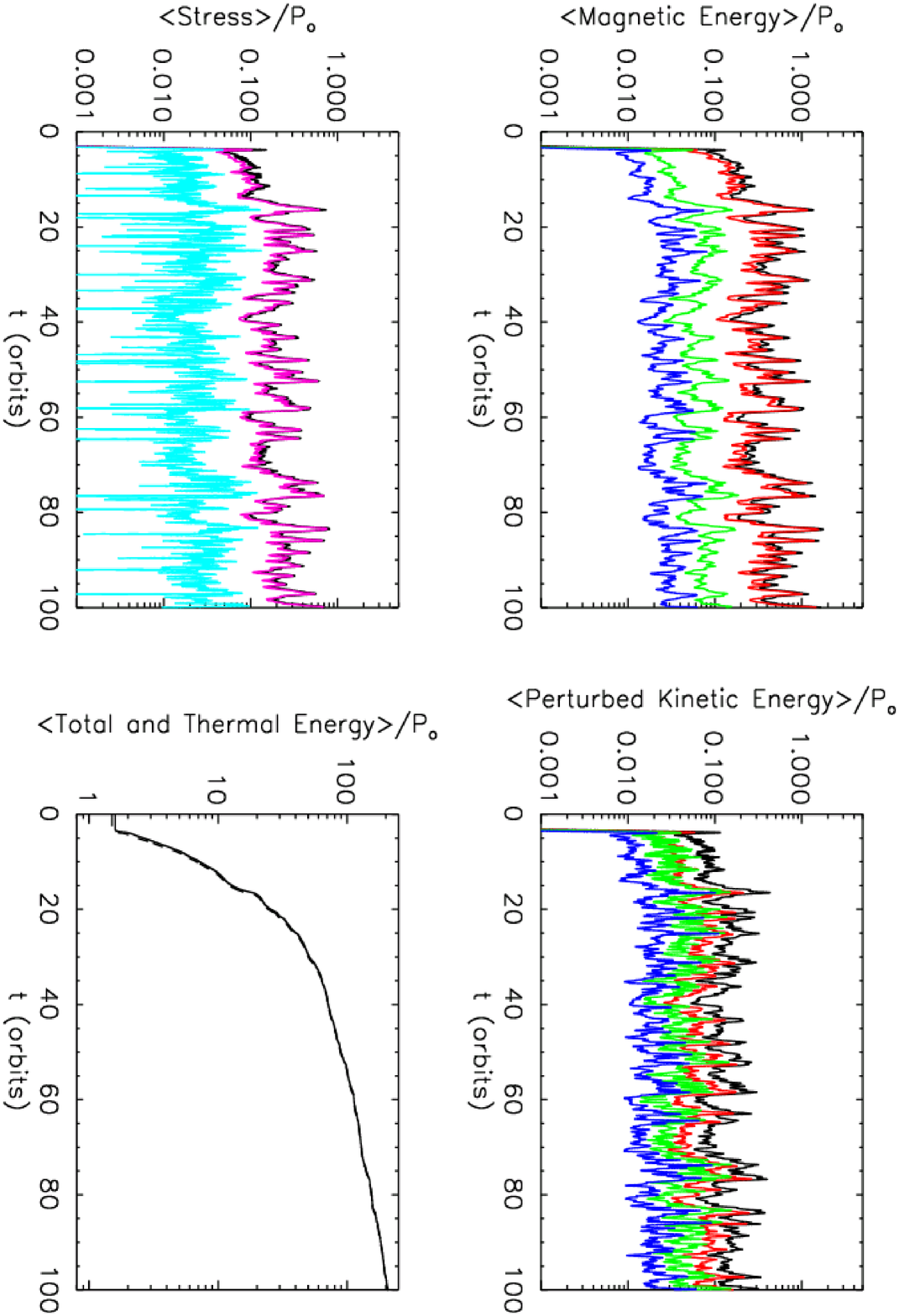}{1in}{90.}{325}{475}{-10}{-100}
\vspace{-1in}
\caption{Volume-averaged energy densities and stresses normalized to the initial gas pressure versus time for the $\none$ simulation.
In the upper two plots, the black line is the total energy density, the green line is the component of the energy density in the $x$ direction, the red line is the $y$ direction component,
and the blue line is the $z$ direction component.
The upper left plot shows the volume-averaged magnetic energy density, the upper right plot shows the perturbed kinetic energy density (i.e., with the shear subtracted off of $v_y$), and the lower left plot is the volume-averaged total stress (black), Maxwell stress (pink), and Reynolds stress (blue).  The lower right plot is the total energy density, including gravitational energy (solid line), and the thermal energy density (dashed line).  The $y$ axes have the same range for all plots except for the total/thermal energy density plot.
\label{turb_n}}
\end{figure}

\begin{figure}
\plotfiddle{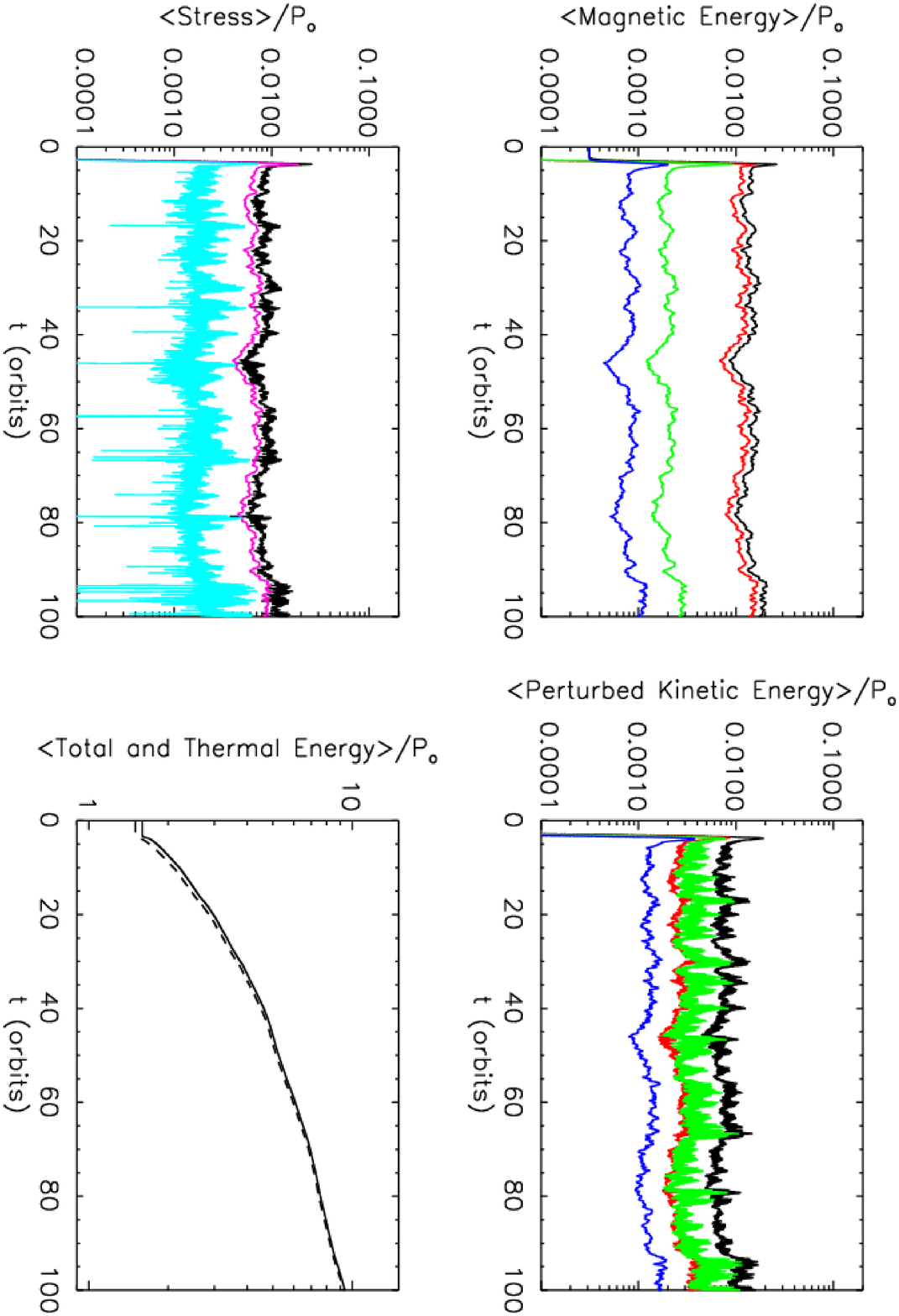}{1in}{90.}{325}{475}{-10}{-100}
\vspace{-1in}
\caption{Volume-averaged energy densities and stresses normalized to the initial gas pressure versus time for the $\zone$ simulation.
In the upper two plots, the black line is the total energy density, the green line is the component of the energy density in the $x$ direction, the red line is the $y$ direction component,
and the blue line is the $z$ direction component.
The upper left plot shows the volume-averaged magnetic energy density, the upper right plot shows the perturbed kinetic energy density (i.e., with the shear subtracted off of $v_y$), and the lower left plot is the volume-averaged total stress (black), Maxwell stress (pink), and Reynolds stress (blue).  The lower right plot is the total energy density, including gravitational energy (solid line), and the thermal energy density (dashed line). The $y$ axes have the same range for all plots except for the total/thermal energy density plot.
\label{turb_z}}
\end{figure}

Figures~\ref{turb_n}~and~\ref{turb_z} show the development of the MRI
and the subsequent evolution of the resulting MHD turbulence for the
fiducial $\none$ and $\zone$ runs respectively.  The MRI saturates
before orbit 5 and the MHD turbulent state lasts for the remainder of
the 100 orbit simulation.  Along with these figures, we list several
time- and volume-averaged quantities from the fiducial runs
in Table~\ref{tbl:sat_char}.  The time average is done from orbits
20 to 100, and the errors are given by one standard deviation over
this period.  Volume-averaged values are indicated by the single-angled
 bracket notation (e.g., $\langle B^2 \rangle$), and time- and volume-
averaged values are denoted by double-angled brackets (e.g., $\langle
\langle B^2 \rangle \rangle$). In both fiducial
runs, the toroidal field magnetic energy dominates with $\langle
B_y^2/2\rangle > \langle B_x^2/2\rangle > \langle B_z^2/2\rangle$.
Examining the components of the kinetic energy and perturbed kinetic
energy, which is $(\rho/2)(v_x^2+\delta v_y^2+v_z^2)$ with $\delta v_y$
given by equation~(\ref{pert_vy}), we find they are closer to each other
in value than are the components of the magnetic energy.  The relative
ordering is similar except that the $x$ kinetic energy is larger than the
perturbed $y$ kinetic energy, $\rho \delta v_y^2/2$, in $\zone$.  Another feature
of note is the greater saturation level and fluctuation amplitude of
the $\none$ run compared to that of $\zone$.  As in past studies, the
Maxwell stress dominates over the Reynolds; the ratio of the Maxwell to
Reynolds stress oscillates between 1 and 10.  Similarly, past studies
have shown a tight correlation between Maxwell (and total) stress and
the magnetic energy density \cite[see, e.g.,][]{black08}.  Here the ratio
of the Maxwell stress to the magnetic energy density is roughly 1/2.
These values and the overall observations are generally consistent with
the results of \cite{haw95}, \cite{haw96}, and \cite{sano04}.

One major difference from past ZEUS simulations is
the evolution of the total ($E+\rho\Phi$) and thermal ($\epsilon$)
energy densities, shown in the lower right plot of
Figs.~\ref{turb_n}~and~\ref{turb_z} for the $\none$ and $\zone$
runs respectively.  Since we evolve an adiabatic equation of state
and there is no cooling term in the energy equation, the total energy
increases with time at a rate given by equation~(\ref{ein}).  The total
energy increases because the free energy of the shearing fluid is being
thermalized by the turbulence, but the shearing box boundary conditions
continuously reinforce that shear.  The stresses at the radial boundaries
therefore constitute a source term.  Equation~(\ref{ein}) also explains
why the total energy reaches a higher value at the end of the simulation
in $\none$ compared to $\zone$.  Since the volume-averaged stress (which
is roughly equal to the stress at the radial boundaries) is higher in
$\none$, the energy injection rate will be larger. These plots also
show that the thermal energy follows the total energy very closely.
That is, the injected energy ends up as thermal energy a short time
later \cite[]{gard05b}.
We will further study the thermalization of injected energy in
\S\ref{energy_fluctuations} and \S\ref{trans_funcs}.

Does the significant increase in thermal energy affect the turbulence
in any way?  This question was examined by \cite{sano04} in an extensive
series of simulations.  They found evidence of a very weak dependence of
the time-averaged Maxwell stress on the gas pressure.  Such an increase
is not apparent from a first look at Figs.~\ref{turb_n} and \ref{turb_z},
but short timescale fluctuations are a dominant feature of these
volume-averaged quantities.  We examined the long term behavior of
the Maxwell stress using time-averaging procedures to smooth away
the fluctuations (which do not appear to change over long timescales).
We found marginal evidence for a weak dependence of the Maxwell stress on
the gas pressure in some, but not all, of the data.  While it is possible
that longer evolution times and a wider exploration of parameter space
could be useful to address this question further, it is clear the stress
has barely changed despite an increase in thermal pressure by a factor
of order 100 in run $\none$.   Thus if there is any dependence of the
stress on the pressure, it is very weak and does not significantly affect
the characteristics of local MRI turbulence.

\begin{figure}
\plotfiddle{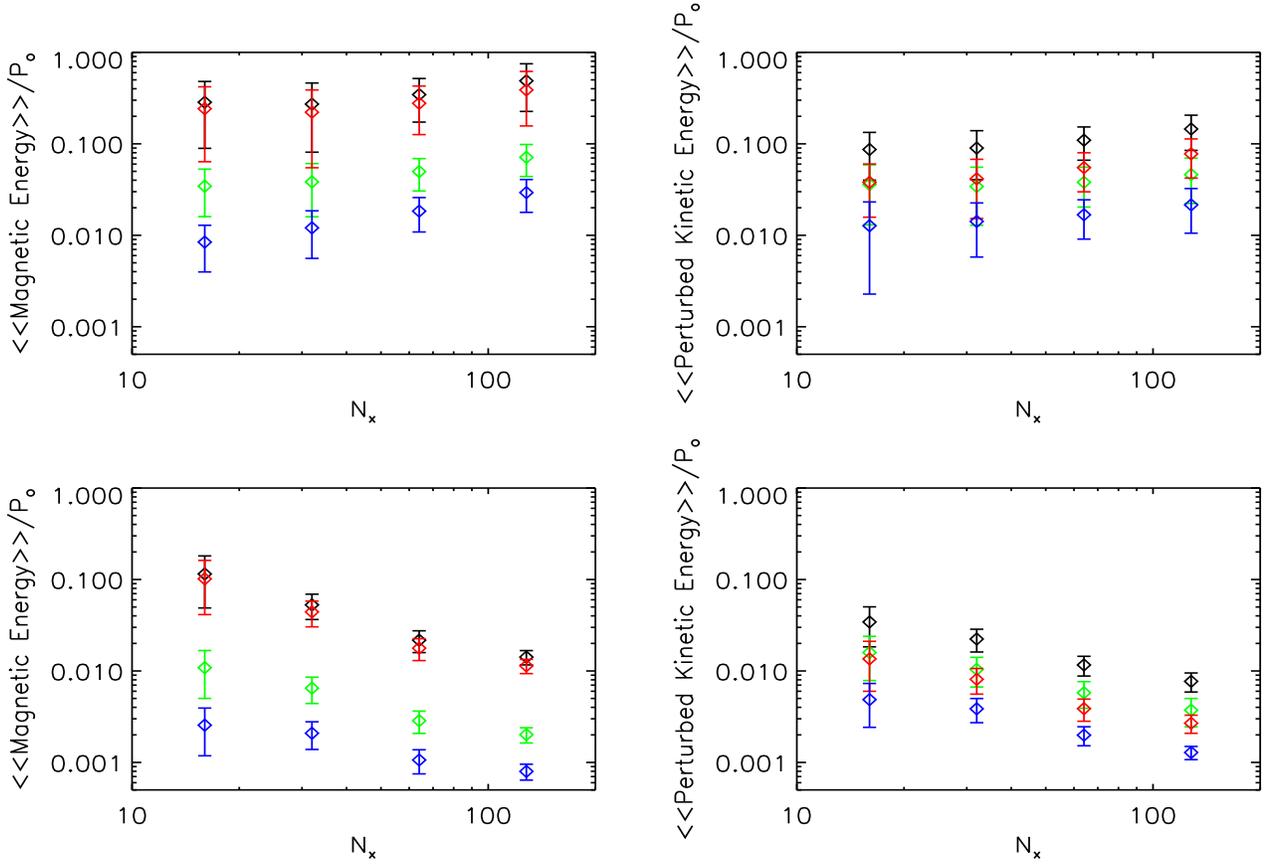}{1in}{90.}{325}{475}{-10}{-100}
\vspace{-1in}
\caption{Time- and volume-averaged energy densities normalized to the initial pressure for various resolutions.
The two upper plots correspond to the net flux simulations, and the two lower plots correspond to the
zero net flux simulations.  The left plots are the averaged magnetic energy densities, and the right plots
are the averaged perturbed kinetic energy densities (i.e., with shear subtracted off of $v_y$).   In all plots, the
black symbols are the total energy density, the green symbols are the $x$ component of the energy density, the
red symbols are the $y$ component, and the blue symbols are the $z$ component.
The time averages are done from orbit 20 to 100, and the error bars indicate one standard deviation
over this period.  
\label{turb_res}}
\end{figure}

We study the effect of resolution through a series of lower resolution
simulations (see Table~\ref{tbl:runs}).  Figure~\ref{turb_res} shows the
time- and volume-averaged magnetic and perturbed kinetic energies as a function of grid
resolution for both the net flux and zero net flux initial conditions.
The time average is calculated from orbits 20 to 100; the error bars
indicate one standard deviation.  For the net flux simulation, there
appears to be a slight trend of increasing energy with resolution,
as observed in \cite{haw95}.  Resolution has a more obvious effect
on the zero net flux initial condition.  The  turbulent energies
{\it decrease} with increasing resolution.  This resolution effect
was previously reported for zero net field initial conditions in other
simulations \cite[]{from07a,pess07} using different numerical algorithms.
With $\at$, the time- and volume-averaged total magnetic energy density
decreases by roughly a factor of two for each factor of two resolution
increase.  The amplitude of the fluctuations in the total magnetic
energy density decreases by roughly a factor of two to four for each
resolution increment.  At all resolutions, the $y$ magnetic energy
density continues to be the largest, followed by the $x$ energy, and
then the $z$ energy.  As was the case for $\none$, $\rho \delta v_y^2/2$
dominates for all net flux simulations, followed by $\rho v_x^2/2$, and
then $\rho v_z^2/2$.  In the zero net flux simulations, the $x$ kinetic
energy density is greater than the perturbed $y$ kinetic energy density.
These components of the perturbed kinetic energy density are close in
value, and it is often the case that the $x$ and $y$ components are
within one standard deviation of each other. The ratio of time- and
volume-averaged Maxwell stress to time- and volume-averaged magnetic
energy density is constant with resolution.  The ratio of time- and
volume-averaged Maxwell stress to time- and volume-averaged Reynolds
stress has a slight increase with resolution in the net flux simulations
and a slight decrease with resolution in the zero net flux simulations.
However, we point out that the observed trends in the ratio of stresses
are subject to considerable uncertainty given the large error bars
calculated for the various quantities.

\subsection{Channel Solution}
\label{channel_solution}

One of the interesting aspects of the vertical field MRI in a shearing
box is that the fastest growing mode leads to axisymmetric
radial streaming motions, dubbed ``channel solutions'' \cite[]{haw92}.
\cite{good94} pointed out that for the vertical field in an
unstratified box, the linear MRI eigenmode is also a nonlinear solution in the
incompressible limit.  They further show that the nonlinear channel solution is itself
unstable to ``parasitic modes.''  These modes require radial and
azimuthal wavelengths larger than the vertical wavelength of the
channel solution and will disrupt the channel flow if the box is
large enough \cite[]{balb98}.

\begin{figure}
\plotfiddle{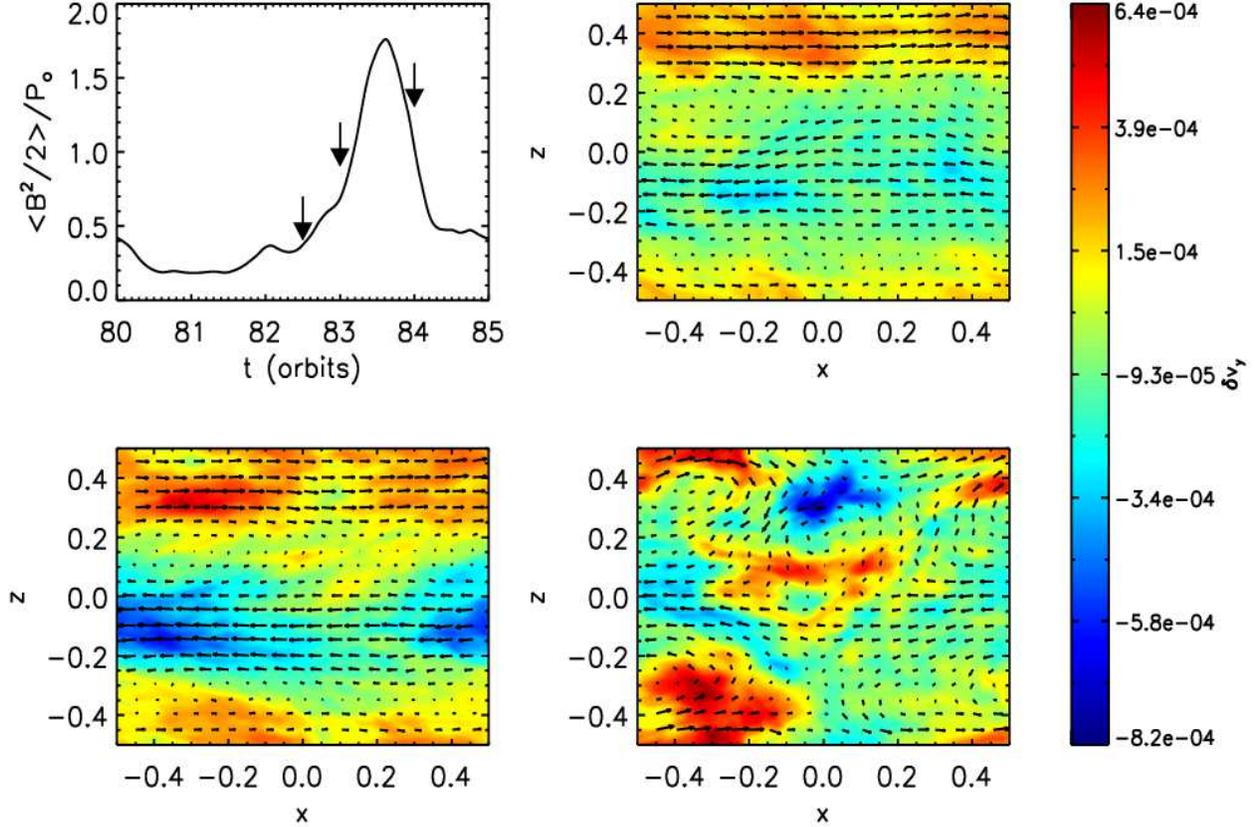}{1in}{90.}{325}{475}{-10}{-100}
\vspace{-1in}
\caption{The development and destruction of a channel flow during the $\none$ simulation.  The upper left plot shows a fluctuation in the volume-averaged magnetic energy density from $t~=~80$~orbits to $t~=~85$~orbits.  The remaining plots show the $y$-averaged perturbed $y$ velocity (colors) and $v_x$ and $v_z$ (vectors). 
 The upper right plot occurs at $t = 82.5$~orbits, the lower left plot occurs at $t = 83$~orbits, and the lower right plot occurs at $t = 84$~orbits.  These times are indicated on the upper left plot by the arrows.  
At $t = 82.5$~orbits, one can see the development of a two-channel flow, in which one channel has $v_x < 0$ and $\delta v_y < 0$, and the other channel has $v_x > 0$ and $\delta v_y > 0$.
 At $t = 83$~orbits, this channel flow is even more developed as the perturbations to the $y$ velocity have
become even stronger and $v_x$ dominates over $v_z$ everywhere.  The development of this channel flow coincides with an increase in volume-averaged magnetic energy density.  By $t = 84$~orbits, the channel flow has been destroyed, coinciding with the decrease in magnetic energy density. 
\label{channel}}
\end{figure}

In the present simulations, the initial vertical field is sufficiently
weak that the fastest growing vertical wavelength is less than the radial
and azimuthal dimensions of the box, and any initial tendency toward
the channel solution at the end of the linear growth phase is quickly
disrupted.  However, we find that the large fluctuations in the magnetic
energy density for $\none$ are a result of recurring channel solutions.\footnote{
The recurrence of the channel solution presumably results from the fact that the net vertical
magnetic field can never be destroyed or removed from the domain, given the periodic
boundary conditions and the strict conservation of $z$ magnetic flux.}
Figure~\ref{channel} shows the azimuthally-averaged velocities at
several times during the amplification and subsequent decay of one such
fluctuation.  The flow organizes itself into a two-channel solution,
which becomes more well-defined as the magnetic energy increases.
The channel solution is eventually destroyed via secondary, parasitic
instabilities \cite[see][]{good94}, which coincides with a decrease in
magnetic energy.  The same channel solution appears during other instances
of large magnetic energy fluctuation in $\none$ and does not appear in
$\zone$. Furthermore, the recurring channel flows appear in the lower
resolution net magnetic flux simulations. As observed previously, the
channel solution and large magnetic energy fluctuations are a property
of simulations with a uniform $B_z$ field \cite[]{sano01}.

Since the channel solution is subject to parasitic modes that depend
on the available wavelengths that can fit in the box, we expect
that this behavior is influenced by the domain aspect ratio employed.
To verify this, we have run several low resolution simulations (labelled 
$\narone-\narsix$, see \S~\ref{method}) using different aspect ratios.
We found that for large enough $L_x$, the intermittent channel modes
no longer occur; this behavior was also observed by \cite{bod08}.
The prominence of intermittent channel flows is a consequence of the
restrictions introduced by the domain size.  However, we use this
property in \S~\ref{energy_fluctuations}, where the large fluctuations
in turbulent energy created by the channel solutions provide a clear
marker of energy injection by the boundaries.  We can then track the
subsequent thermalization of that energy.

\subsection{Energy Power Spectra}
\label{energy_spec}

The nature of MRI-driven MHD turbulence can be characterized in part
by the power spectrum of kinetic and magnetic energies.  To obtain such
power spectra, we do a full 3D Fourier transform on the simulation data
employing the procedures outlined in \cite{haw95} to account for the
shearing-periodic boundaries.  Briefly, the shearing periodic boundary
conditions in the $x$ direction allow the domain to be strictly periodic
in the $x$ direction only at certain times, called periodic points $t_n$
(described in \S\ref{method}).   To perform a standard fast Fourier
transform (FFT) at some time $t$ that is not equal to $t_n$, we transform
the data into a frame where the $x$ boundaries are strictly periodic.
We then calculate the FFT in this frame and remap to the original frame.

The turbulent magnetic, kinetic, and perturbed kinetic energy densities
in Fourier space are defined as

\begin{equation}
\label{mag_fourier}
\frac{1}{2}|\bfour|^2 \equiv \frac{1}{2}\left[|\widetilde{B_x}({\bm k})|^2+|\widetilde{B_y}({\bm k})|^2+|\widetilde{B_z}({\bm k})|^2\right],
\end{equation}

\begin{equation}
\label{ke_fourier}
\frac{1}{2}|\vfour|^2 \equiv \frac{1}{2}\left[|\widetilde{\sqrt{\rho}v_x({\bm k})}|^2+|\widetilde{\sqrt{\rho}v_y({\bm k})}|^2+|\widetilde{\sqrt{\rho}v_z}({\bm k})|^2\right],
\end{equation} 

\begin{equation}
\label{pke_fourier}
\frac{1}{2}|\vpfour|^2 \equiv \frac{1}{2}\left[|\widetilde{\sqrt{\rho}v_x}({\bm k})|^2+|\widetilde{\sqrt{\rho}\delta v_y({\bm k})}|^2+|\widetilde{\sqrt{\rho}v_z({\bm k})}|^2\right],
\end{equation} 

\noindent
where $\widetilde{f}$ means the Fourier transform of $f$ defined by

\begin{equation}
\label{four_trans}
\widetilde{f({\bm k})} = \int \int \int f({\bm x}) \fterm.
\end{equation}

\noindent Note that for the kinetic energies, we include the density along
with the velocity when calculating the Fourier transform, resulting in
the appearance of $\sqrt{\rho}$ in the above equations.  To obtain these
quantities as a function of length scale and to improve statistics, we
average our data over shells of constant $k = |{\bm k}|$.  For further
improvement of statistics, we average each of these terms over 161 frames
(i.e., from orbit 20 to 100 in increments of 0.5 orbits).

\begin{figure}
\plotfiddle{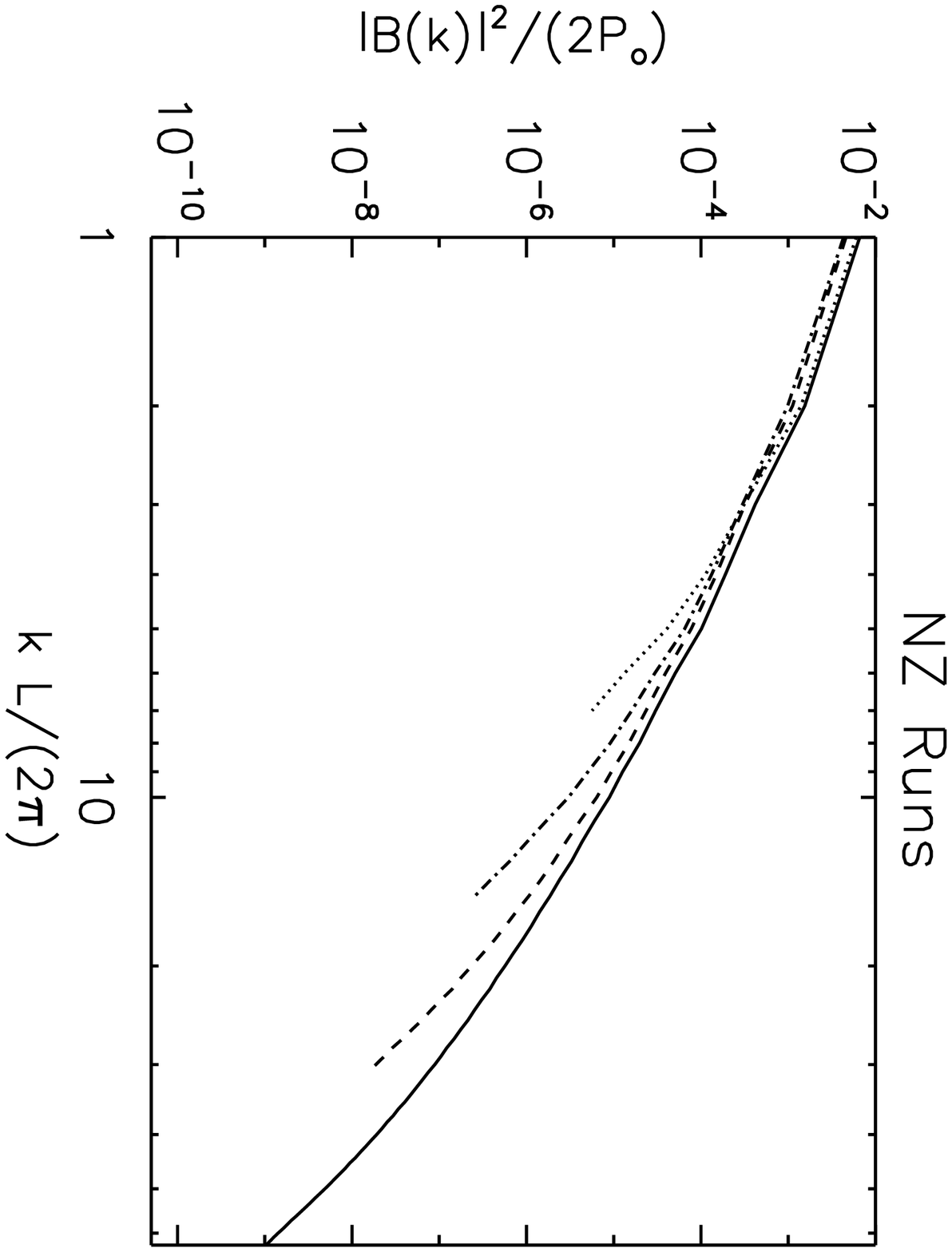}{0in}{90.}{125}{175}{20}{-100}
\vspace{-2.1in}
\plotfiddle{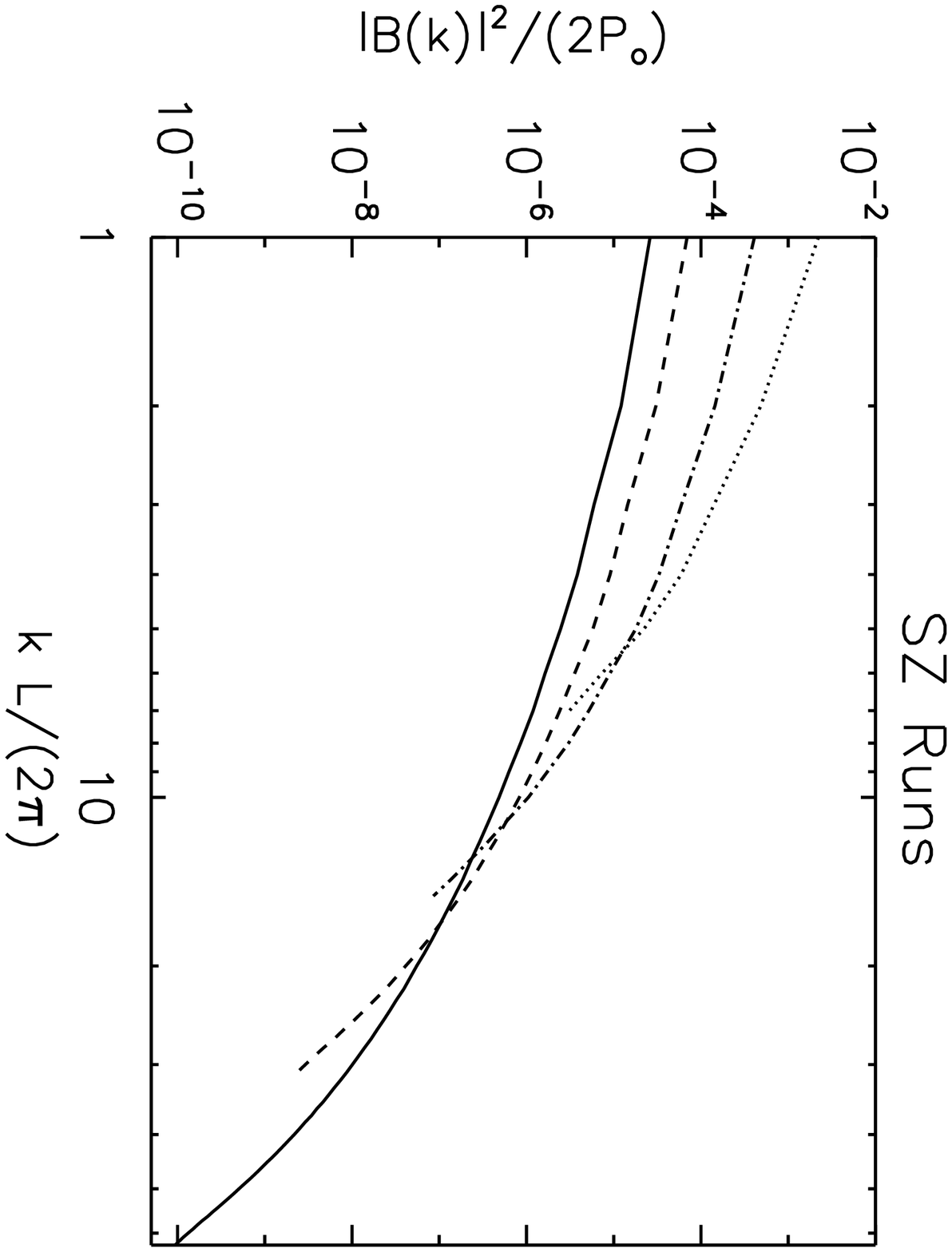}{0in}{90.}{125}{175}{255}{-100}
\plotfiddle{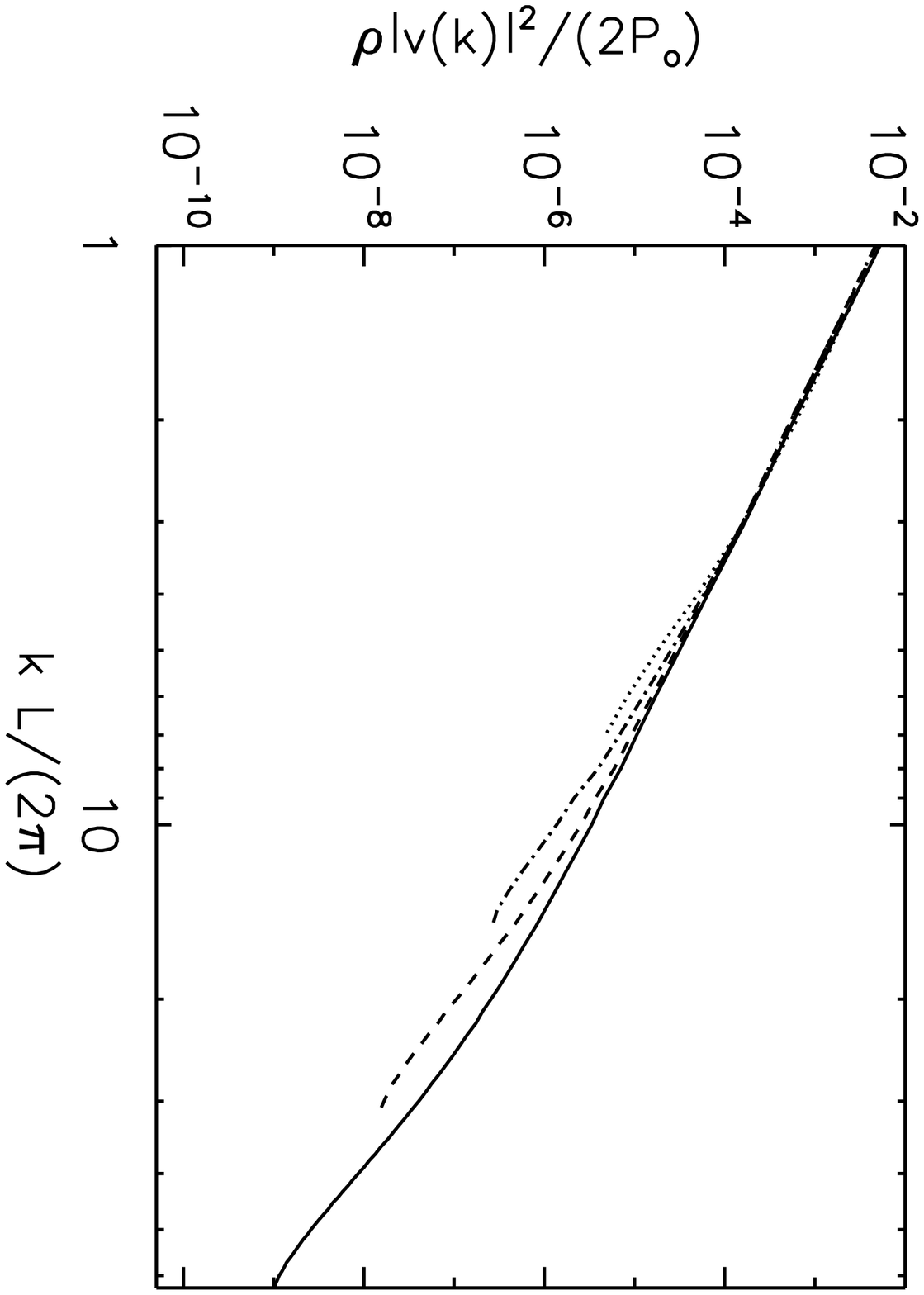}{0in}{90.}{125}{175}{20}{-100}
\vspace{-2.1in}
\plotfiddle{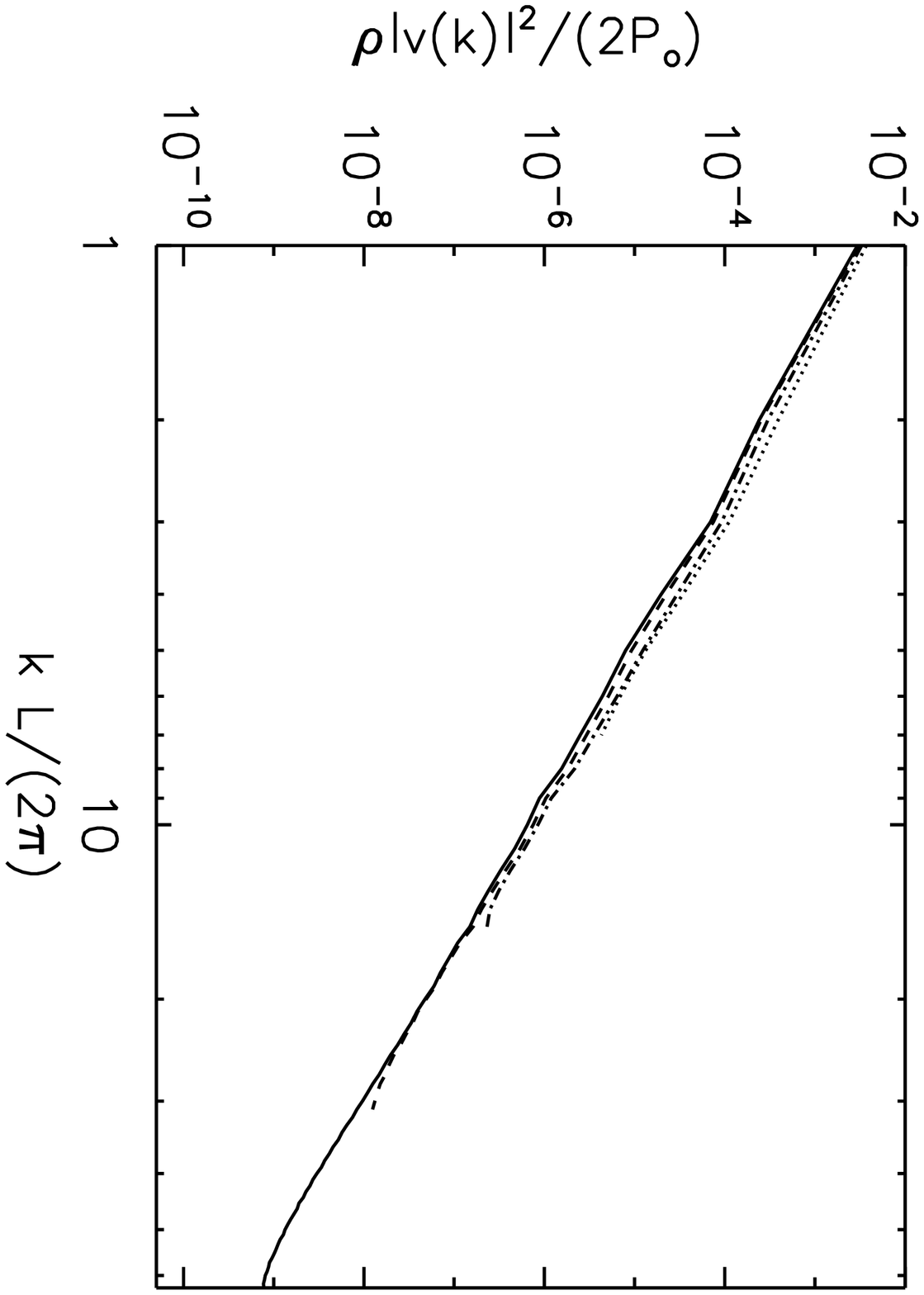}{0in}{90.}{125}{175}{255}{-100}
\plotfiddle{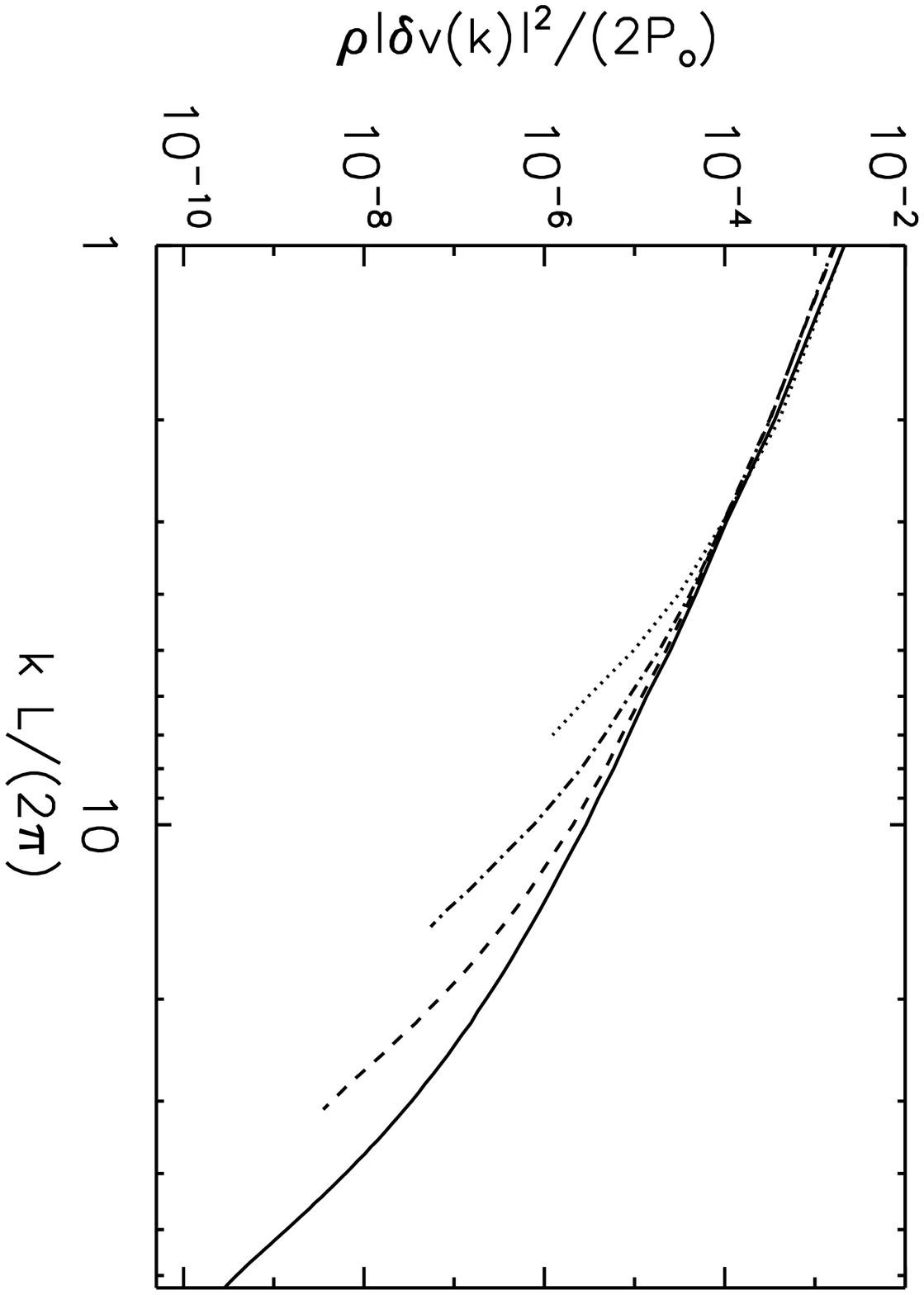}{0in}{90.}{125}{175}{20}{-100}
\vspace{-2.1in}
\plotfiddle{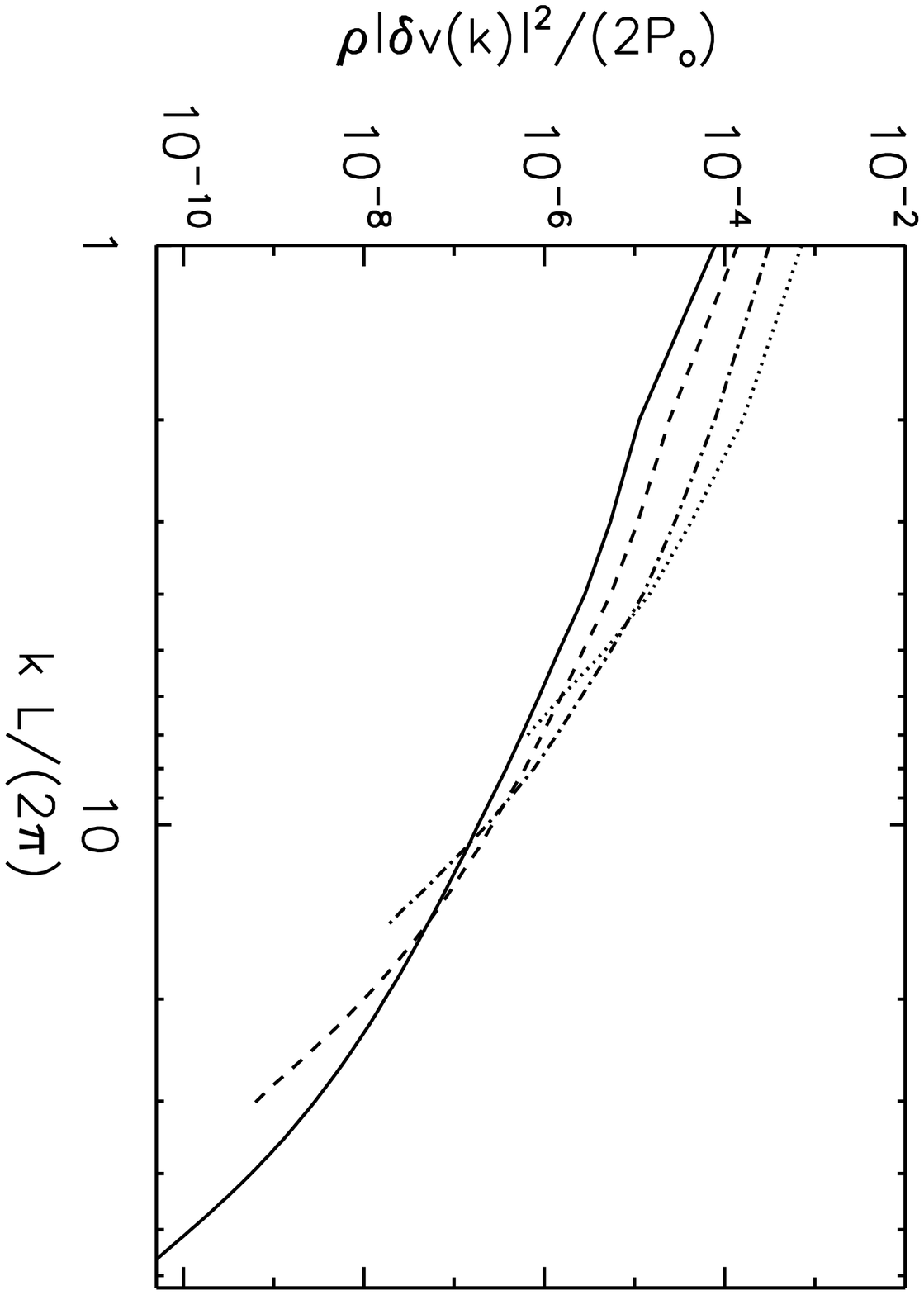}{0in}{90.}{125}{175}{255}{-100}
\caption{Spatial power spectra of various energy densities in the saturated state of the standard net flux (left panels) and zero net flux simulations (right panels).  The spectra were obtained via an average over 161 frames in the saturated state and an average over shells of constant modulus $|{\bm k}|$.  In each column, the first plot shows magnetic energy density, the second shows kinetic energy density, and the third shows perturbed kinetic energy density (as defined in the text).  The effect of resolution is shown in each individual plot; the dotted line corresponds to the resolution with $N_x = 16$, the dot-dashed line corresponds to $N_x = 32$, the dashed line corresponds to $N_x = 64$, and the solid line corresponds to $N_x = 128$.  
All energy densities have been normalized to the initial gas pressure and are plotted against a dimensionless wave number ($L$ is the length of the smallest dimension of the box).
\label{power_spec}}
\end{figure}

Figure~\ref{power_spec} shows the power spectra
of these energy densities for the net flux and zero net flux runs.  
The figure shows resolution effects as different lines in each plot.
In all cases, the largest scales account for most of the energy.  The
general shape of the energy power spectra agrees with previous
studies \cite[e.g.,][]{haw95,from07a}.  For the net flux simulations,
the magnetic energy dominates over the kinetic and perturbed kinetic
energies at all scales, independent of resolution.  As the resolution
is increased, the power spectra extend to higher $k$, but the general
shape remains constant.   
At some values for $k$, the uncertainty in
energy (not plotted), represented by one temporal standard deviation around the mean,
 is large enough to overlap with other energy components, making it difficult
to conclusively say which energy dominates at these particular scales.

We calculated a power law index in Fourier space for each energy density
and at each resolution.  This slope was determined by a linear fit to
the energy densities in log space from $k L/(2\pi) = 1$ to the maximum
scale for the given resolution.  There is some uncertainty in this measurement
because the power spectra are not strictly linear in log space (see Fig.~\ref{power_spec}).
In $\none$, the energy density is
proportional to $[k L/(2\pi)]^n$ with $n \approx -4$ for every energy
density.  This index is approximately constant with resolution, but
there is evidence that $n$ becomes more negative at higher resolutions.
In determining an error in the value of $n$, we found that this error is
often dominant.  Thus, such a resolution dependence is somewhat tentative.

There is a noticeable resolution dependence in the zero net
flux simulations.  First, as resolution is increased, the magnetic
energy density decreases at all scales.  This effect was discussed in
\S~\ref{sat_characteristics}; the power spectra are consistent with the
power spectrum analysis of \cite{from07a}.  The same resolution dependence
is observed for the perturbed kinetic energy density.  The magnetic
energy density at small $k$ decreases faster with resolution than does
the perturbed kinetic energy density. The total kinetic energy density (i.e.,
including shear) remains constant with resolution, which simply results
from the fact that the shear velocity, which dominates the kinetic energy,
is constant with resolution.  The uncertainty in each energy component
appears to be smaller than in the net flux simulations. However, there
are still some values of $k$ at which the calculated errors overlap.

We calculated a power law index in Fourier space for each energy density
and resolution for the zero net flux simulations.  The procedure we
used was the same as for the net flux simulations.  For the kinetic and
perturbed kinetic energy densities, we found that $n$ lies between -3.5
and -4, whereas for the magnetic energy density, $n$ lies between -3
and -3.5.  There does not appear to be any resolution dependence in $n$
for the magnetic energy density, but there is a tentative decrease in $n$
(similar to the net flux case) with increasing resolution for the kinetic
and perturbed kinetic energy densities.

\clearpage
\begin{deluxetable}{c|cc}
\tabletypesize{\scriptsize}
\tablewidth{0pc}
\tablecaption{Saturation Characteristics \label{tbl:sat_char}}
\tablehead{
\colhead{Quantity}&
\colhead{$\none$}&
\colhead{$\zone$}    }
\startdata
$\maxwell/P_o$ & 0.216 $\pm$ 0.116 & $(6.55\pm1.15) \times 10^{-3}$ \\
$\reynolds/P_o$ & 0.028 $\pm$ 0.019 & $(1.91\pm0.76) \times 10^{-3}$ \\
$\magnetic/P_o$ & 0.488 $\pm$ 0.262 & 0.014 $\pm$ 0.003 \\
$\magx/P_o$ & 0.071 $\pm$ 0.027 & $(2.01\pm0.38) \times 10^{-3}$ \\
$\magy/P_o$ & 0.388 $\pm$ 0.231 & 0.011 $\pm$ 0.002 \\
$\magz/P_o$ & 0.029 $\pm$ 0.011 & $(7.98\pm1.57) \times 10^{-4}$ \\
$\pertkinetic/P_o$ & 0.145 $\pm$ 0.060 & $(7.69\pm1.81) \times 10^{-3}$ \\
$\kinx/P_o$ & 0.046 $\pm$ 0.024 & $(3.73\pm1.27) \times 10^{-3}$ \\
$\pertkiny/P_o$ & 0.078 $\pm$ 0.035 & $(2.68\pm0.60) \times 10^{-3}$ \\
$\kinz/P_o$ & 0.021 $\pm$ 0.011 & $(1.28\pm0.21) \times 10^{-3}$ \\
$\maxwell/\reynolds$ & 7.60 $\pm$ 6.47 & 3.43 $\pm$ 1.49 \\
$\maxwell/\magnetic$ & 0.443 $\pm$ 0.336 & 0.462 $\pm$ 0.116 \\
\enddata
\end{deluxetable}
\clearpage

\section{Energy Fluctuations}
\label{energy_fluctuations}

$\at$ evolves the equation for total energy, the volume-average
of which will change only due to the Maxwell and Reynold stresses
at the radial boundaries (equation (\ref{ein})).  As was discussed in
\S\ref{general_properties}, the individual volume-averaged magnetic and
kinetic energies are highly variable throughout the evolution as energy is continuously
transferred between magnetic, kinetic and thermal components.  We can
study these energy flow processes by tracking the energy injected at the boundaries as
it is subsequently thermalized in the turbulence.  For this purpose, the
existence of the recurring channel solution in the net magnetic field
simulation is very useful; the sudden increase in stress provides a
clear injection of energy that can be traced using several diagnostics.
Having developed these diagnostics we can then apply them to the zero
net magnetic flux simulations.  Finally,  to gain additional insight
into dissipation in the turbulence, we conduct an experiment in which
the shear flow and gravity terms have been removed.

\subsection{Sustained Turbulence}
\label{sustained}

The total energy density, including the gravitational potential energy density,
is defined as

\begin{equation}
\label{total_energy}
\etot = E + \rho \Phi = \epsilon + \frac{1}{2} \rho v^2 + \frac{1}{2} B^2 + \rho \Phi
\end{equation}

\noindent
where $\Phi$ is given in \S\ref{method}.  Averaging 
equation~(\ref{total_energy}) 
over the entire domain, taking the time derivative, and
rearranging the terms, we obtain,

\begin{equation}
\label{dtedt}
\td = \ein - \kd - \md - \gd .
\end{equation}

\noindent
where $\ein \equiv \partial \langle \etot\rangle/\partial t$ is the energy
injection rate due to stress at the boundaries (see equation~(\ref{ein})),
$\td \equiv \partial \langle \epsilon\rangle /\partial t$ is the rate of
change of thermal energy density, $\kd \equiv \partial \langle \frac{1}{2}\rho
v^2 \rangle / \partial t$ is the rate of change of the kinetic energy
density, $\md \equiv \partial \langle \frac{1}{2} B^2\rangle /
\partial t$ is the rate of change of the magnetic energy density,
and $\gd \equiv \partial \langle \rho \Phi\rangle / \partial t$
is the time derivative of the tidal potential energy density.
The brackets indicate a 
volume-average over the simulation domain.  $\gd$ is the change
in a fluid element's gravitational energy as it moves within the domain.
We expect the contribution of the tidal potential
term to be insignificant, an expectation borne out by direct computation.
We will ignore this term in most of the subsequent discussion.  The stress
terms at the radial boundaries are generally positive, which means
energy is being injected into the box via the work done by this stress
($\ein >$ 0).  The remaining terms in equation~(\ref{dtedt}) can be
either positive or negative.

\begin{figure}
\begin{center}
\plotfiddle{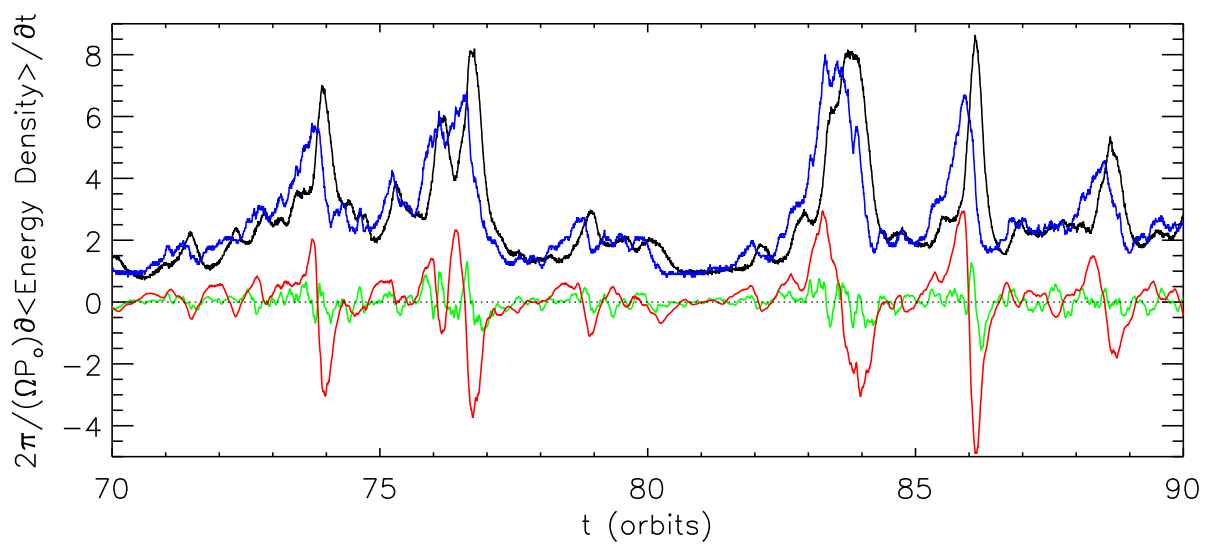}{1in}{90.}{200}{350}{-100}{-100}
\vspace{-1in}
\end{center}
\caption{Time derivative of various volume-averaged energy densities for a 20 orbit period in the $\none$ simulation.
The time derivative of the energy densities have been multiplied by an orbital time over the initial gas pressure.
The dark blue line is the energy injection rate, $\ein$, the black line is the thermal energy density derivative, $\td$, the green line is the kinetic energy density derivative, $\kd$, and the red line is the magnetic energy density derivative, $\md$.  The dotted line indicates zero.
\label{dedt_n}}
\end{figure}

The lower right plot in Figure~\ref{turb_n} shows that the thermal
energy density closely follows the total energy density, but with
a short time delay.  This can be better seen in Figure~\ref{dedt_n},
which shows the individual terms from equation~(\ref{dtedt})
for a 20 orbit period in the $\none$ simulation.  There is a clear time
delay of less than one orbit between significant changes in the energy
injection rate and the thermal energy derivative, suggesting a comparable delay before
the injected energy is thermalized, a property noted in \cite{sano01}
as well as in \cite{gard05b}.  These features in the energy derivatives
result from the creation and destruction of channel flows. During this time interval, the magnetic
and kinetic energies are also changing. By examining the maxima in
the thermal energy derivative and the corresponding features in the kinetic
and magnetic energy derivatives, it appears that the magnetic energy
dissipation dominates the thermalization process.

It is useful to define a temporal correlation function for the various
energy components by writing

\begin{equation}
\label{corr_func}
C_{AB} \equiv \left\{ \begin{array}{ll}
\frac{\displaystyle\frac{1}{N-|L|}\sum_{i=0}^{N-|L|-1} A_{i+|L|} B_i}{\displaystyle\frac{1}{N}\sum_{i=0}^{N-1} A_i} & \quad
\mbox{if $L <$ 0} \\
\frac{\displaystyle\frac{1}{N-|L|}\sum_{i=0}^{N-L-1} A_i B_{i+L}}{\displaystyle\frac{1}{N}\sum_{i=0}^{N-1} A_i} & \quad
\mbox{if $L \ge$ 0}
\end{array} \right.
\end{equation}

\noindent 
where $A$ and $B$ are two time-series datasets $N$ elements
in length.  The quantity $L$ is the number of elements over which to
shift $A$ and $B$ with respect to each other to calculate the correlation
coefficient.  We apply equation~(\ref{corr_func}) to the energy rates
by setting $A$ = $\td$, and $B$ = $\kd$, $\md$, or $\ein$.  This allows
us to correlate the energy injection rate and the change in kinetic and
magnetic energies against the change in thermal energy over certain timescales. Since
$\td >$ 0, if the correlation between $\td$ and $\kd$ (or $\md$) is
negative, then kinetic energy (or magnetic energy) must be decreasing,
and a strong negative correlation would suggest that kinetic energy
(or magnetic energy) is being thermalized. 

\begin{figure}
\plotfiddle{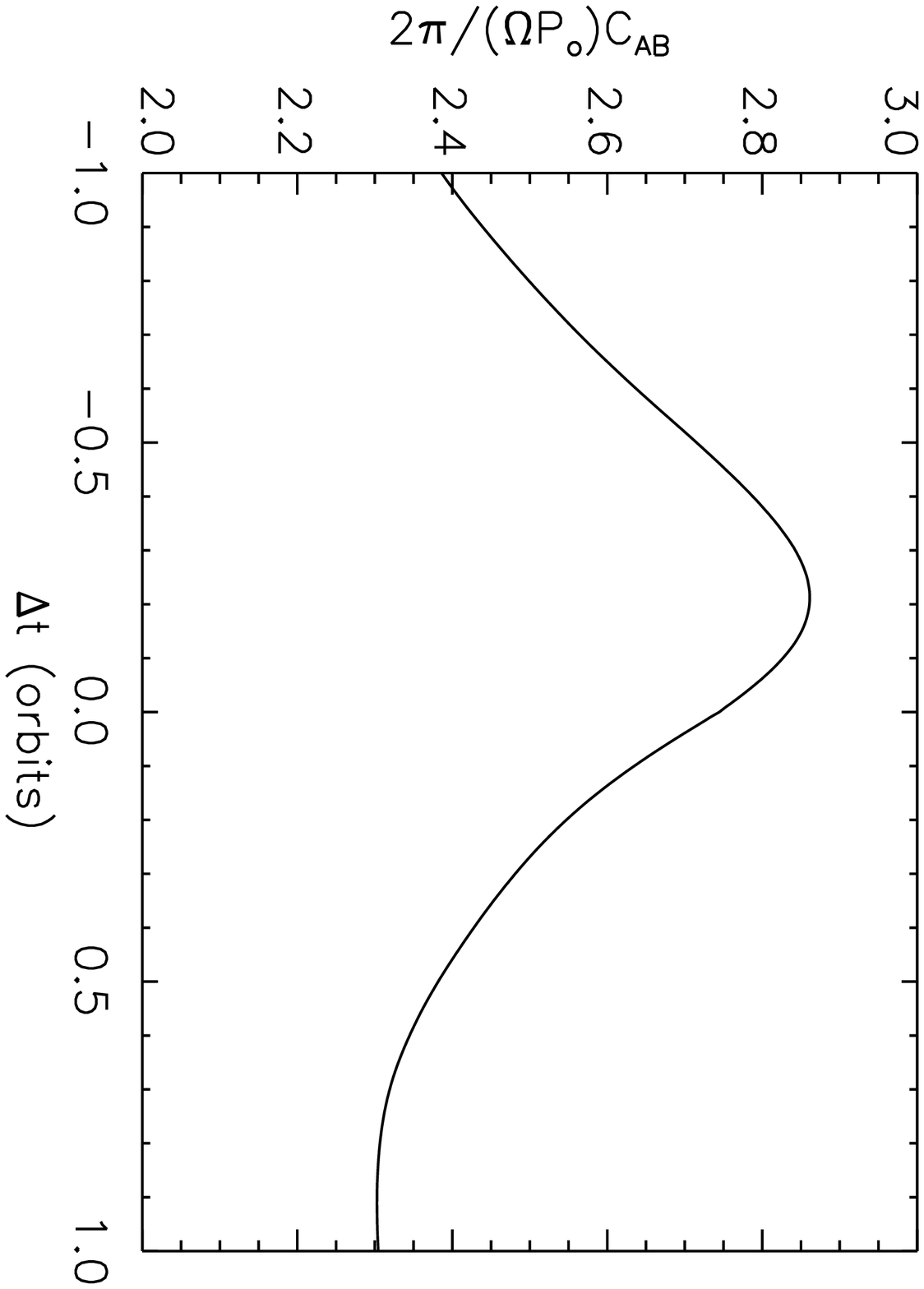}{1in}{90.}{200}{250}{-30}{-100}
\vspace{-3.12in}
\plotfiddle{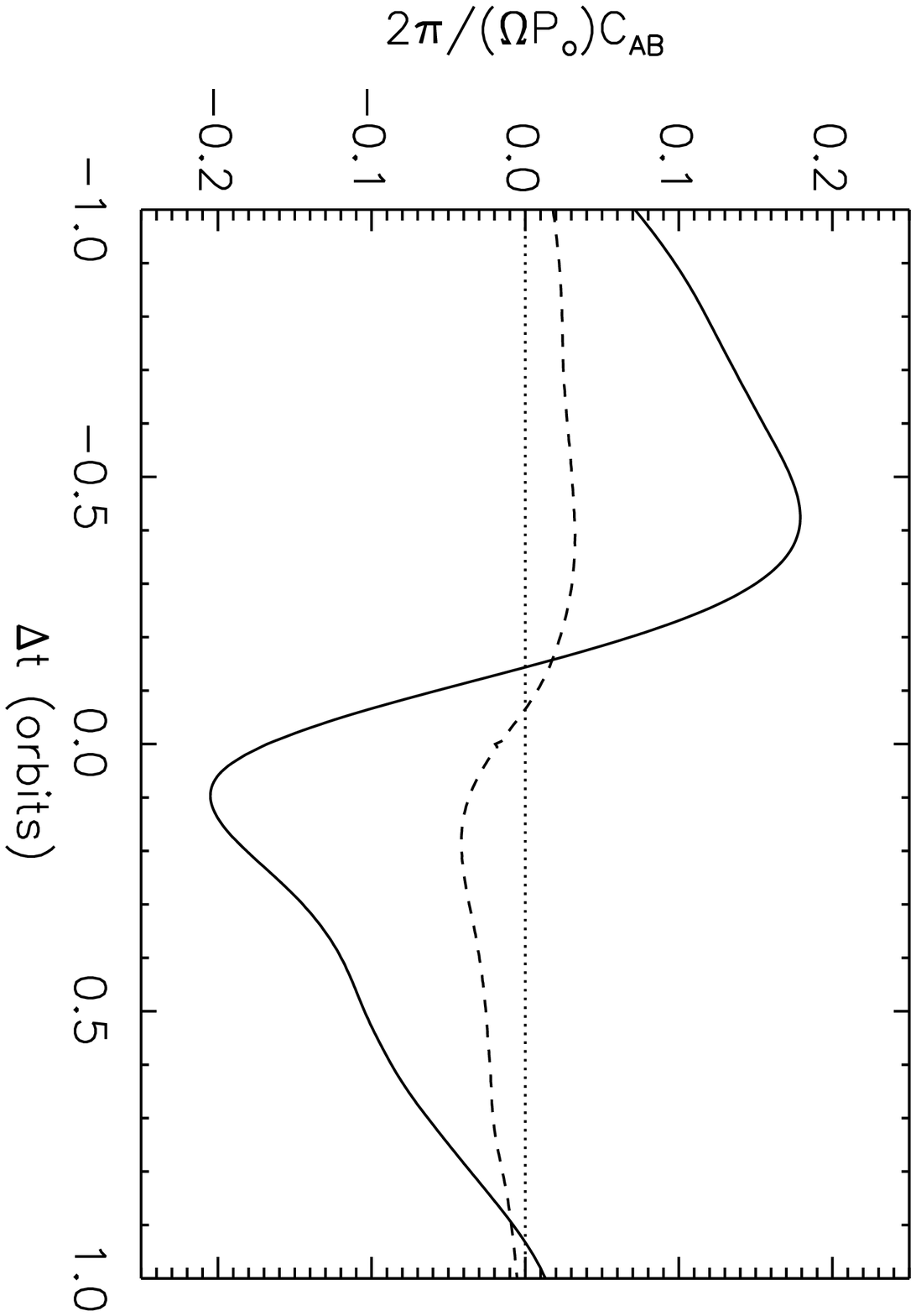}{0in}{90.}{200}{250}{235}{-100}
\caption{Correlation coefficients calculated over orbits 20 to 100 in the $\none$ simulation.
The plot on the left was calculated by correlating the energy injection rate, $\ein$, against the thermal energy time derivative, $\td$.
The $x$-axis is the correlation length in time, and the $y$-axis is the coefficient multiplied by an orbital period over the initial gas pressure.
The plot on the right was calculated by correlating the magnetic energy derivative, $\md$, (solid line) and
kinetic energy derivative, $\kd$, (dashed line) against the thermal energy derivative. The dotted line indicates $C_{AB} = 0$. Note that
the two plots have different $y$ scales.
\label{corr_n}}
\end{figure}

Figure~\ref{corr_n} is the correlation function for $B = \ein, \kd,$ and
$\md$ calculated over orbits 20 to 100.
The $x$-axis is the correlation timescale in units of orbits.  
 We only look at correlation times of $\lesssim$ 1 orbit as the degree to which
the thermal energy evolution follows that of the total energy (see Figures~\ref{turb_n}
and \ref{turb_z}) indicate that thermalization happens over that timescale.  To examine
the correlation function on longer timescales would be misleading since peaks in the function
would suggest a correlation between two events that are not causally related (e.g., the injection
of energy for one channel event being correlated with the thermalization of energy for another channel
event). The left plot of the figure shows that $\ein$ is strongly correlated with
$\td$ on a timescale of $\Delta t \sim$ -0.2 orbits.  This correlation is
exactly what we observed in Fig.~\ref{dedt_n}.  The energy injected by
the stress at the boundaries ends up as heat less than one orbit later.
The negative sign on this value of $\Delta t$ simply means that the
injection happens before the thermalization.   In the right plot,
both $\kd$ and $\md$ are negatively correlated with $\td$ suggesting that
magnetic and kinetic energy are being thermalized.  The stronger magnetic correlation further
suggests that magnetic dissipation contributes more to thermalization than
kinetic dissipation.  The positive correlation between $\kd$ and $\md$
against $\td$ at negative $\Delta t$ values is a result of the magnetic
and kinetic energies increasing along with the energy injection into
the box.  That is, the stress at the boundaries increases the magnetic
and kinetic energies which are dissipated a short time later.

An interesting feature is evident in Fig.~\ref{corr_n}: the negative
peak in the magnetic and kinetic correlation functions occur for $\Delta
t$ slightly greater than zero.  Similarly, in Fig.~\ref{dedt_n} one
can see that peaks in the magnetic and kinetic energy derivatives are
offset with respect to the energy injection and thermalization peaks.
For example, the maximum rate for magnetic energy loss occurs after
the maximum rate for thermal energy gain.  Of course, these are plots of
the time derivative of the energy, so a peak simply indicates where the
second derivative is zero.  The magnetic energy is both losing energy
to dissipation while gaining energy from the shear at the boundaries.
When the energy injection rate peaks decline, the thermalization rate
is still growing and the magnetic energy rate also peaks and begins
to decline.  Similarly, the slope of the magnetic energy loss rate will
change sign after the thermalization rate has peaked and when the energy
injection rate is no longer itself in decline.

As a test, we performed this correlation analysis on
the lower resolution net-flux simulations and find that energy injection
precedes thermalization by $\sim$ 0.2 orbits, independent of resolution.
Furthermore, magnetic dissipation dominates over kinetic dissipation for all
net flux simulations.

\begin{figure}
\plotfiddle{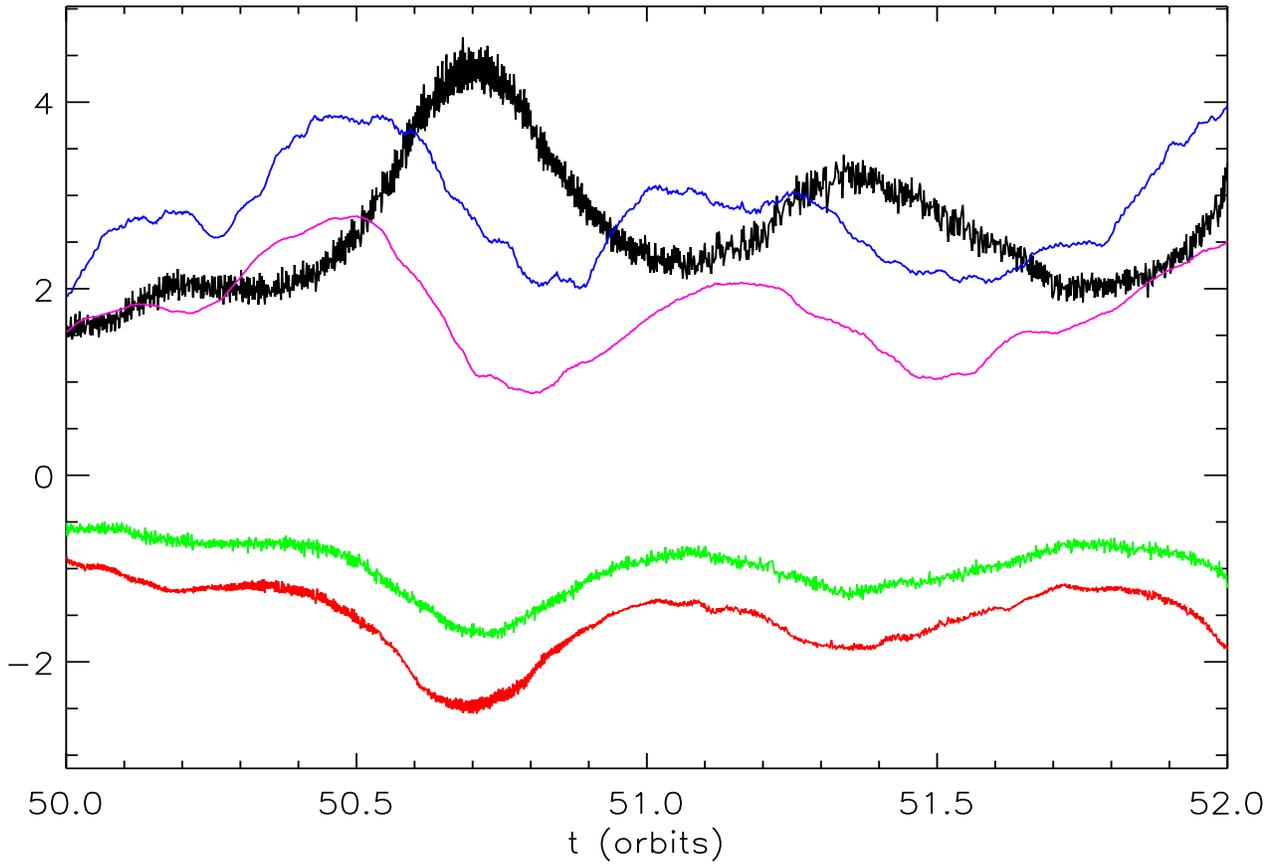}{1in}{90.}{325}{475}{-10}{-100}
\vspace{-1in}
\caption{Various terms in the volume-averaged magnetic, kinetic, and thermal energy density evolution equations over a two orbit period of $\none$. 
 The energy terms are $\td$ (black),
$\ein$ (blue), -$\qk$ (green), -$\qm$ (red), and the volume-averaged transfer rate from kinetic to magnetic energy (pink). All of these terms
are defined in the text and have been multiplied by an orbital period over the initial gas pressure.
\label{energies_n}}
\end{figure}

The analysis so far has only examined the rate of change in the energy
terms, not specifically how they change.  For example, does a ``dip"
in $\md$ correspond to direct thermalization of magnetic energy, or is
there a transfer of energy from magnetic to kinetic?  To examine the
energy flow in more detail, we focus on orbits 50 to 52 in $\none$, for which we ran the $\none$ simulation at 
high temporal resolution.  This high time resolution allows us to resolve short timescale features, but also
generates many large data files.  Therefore, we restrict this part of the analysis to the two orbit period mentioned above.
Consider the evolution equation for the volume-averaged kinetic energy
given by

\begin{eqnarray}
\label{ke_eqn}
\kd & = & -\left\langle \del \cdot \left[{\bm v}\left(\frac{1}{2}\rho v^2+\frac{1}{2}B^2 + P + \rho \Phi \right) - {\bm B} ({\bm v} \cdot {\bm B})\right]\right\rangle \nonumber \\
& & + \left\langle\left(P + \frac{1}{2} B^2\right)\del \cdot {\bm v}\right\rangle-\left\langle{\bm B} \cdot ({\bm B} \cdot \del {\bm v})\right\rangle - \gd - \qk,
\end{eqnarray}

\noindent
where $\qk$ is the volume-averaged (numerical) kinetic energy dissipation rate. 
The evolution equation for the volume-averaged magnetic energy is given by

\begin{equation}
\label{be_eqn}
\md = - \left\langle \del \cdot \left(\frac{1}{2}B^2{\bm v} \right)\right\rangle - \left\langle \frac{1}{2} B^2 \del \cdot {\bm v}\right\rangle + \left\langle{\bm B} \cdot ({\bm B} \cdot \del {\bm v})\right\rangle- \qm
\end{equation} 

\noindent 
where $\qm$ is the volume-averaged (numerical) magnetic energy
dissipation rate.  We have calculated each term in these equations over
the two orbit period and find that the dominant terms are $-\langle\del
\cdot (\frac{1}{2}\rho v^2){\bm v}\rangle$, $\langle\del \cdot [{\bm
B}({\bm v} \cdot {\bm B})]\rangle$, $\langle{\bm B} \cdot ({\bm B} \cdot
\del {\bm v})\rangle$, $\qk$, and $\qm$.  $\qk$ and $\qm$ are what remain
after calculating all other terms in the energy equations at a particular
instant in time. Calculating the volume-averages
of the first two terms yields the radial boundary Reynolds and Maxwell
stresses in Equation~(\ref{ein}) \cite[]{haw95}, namely the energy
injection rate by the shearing periodic boundaries.  The third of
the dominant terms is the transfer rate of kinetic to magnetic energy via
field line stretching.  Figure~\ref{energies_n} plots the time-history of
this term (pink line) along with $\td$ (black line),  the energy injection
rate $\ein$ (blue line),  and $-\qk$ and $-\qm$ (green and red lines, respectively).
As energy is injected into the grid, a significant fraction of this
energy is transferred to the magnetic field via field line stretching,
presumably through the shear flow.  Thermalization follows 0.2 orbits
later and is marked by increases in the
absolute value of $\qk$ and $\qm$, with $|\qm| > |\qk|$.  The ratio of
kinetic to magnetic dissipation is approximately constant in time over
this period, with $\qk/\qm \approx 0.6$.  This suggests that the details
of the thermalization do not vary with intermittent increases in $\ein$
that occur when the fluid experiences a channel flow.

\begin{figure}
\plotfiddle{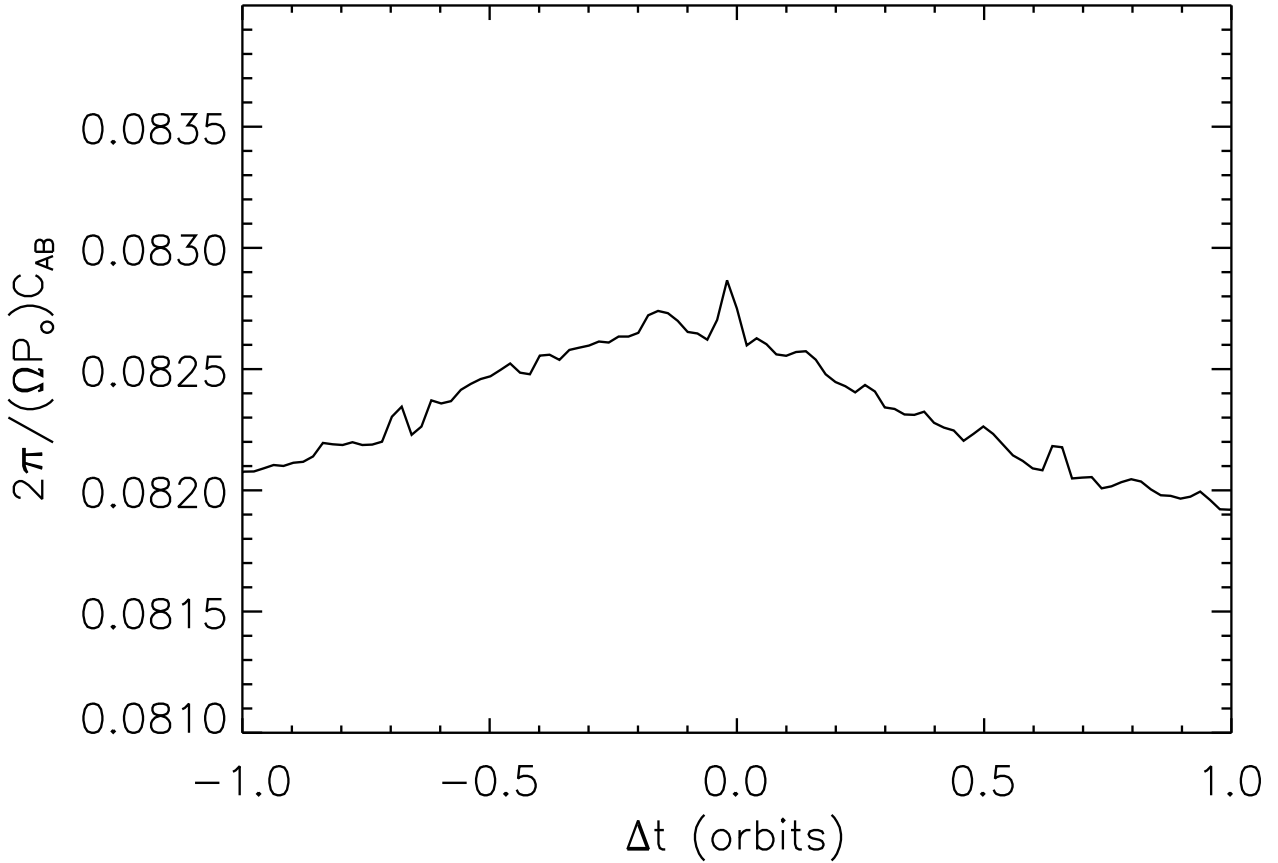}{1in}{90.}{325}{475}{-10}{-100}
\vspace{-1in}
\caption{Correlation coefficient calculated over the saturated state of the $\zone$ simulation.
The coefficient was calculated by correlating the energy injection rate against the thermal energy density derivative.
The $x$-axis is the correlation length in time, and the $y$-axis is the coefficient multiplied by an orbital period over the initial gas pressure.
The narrow peaks in the curve correspond to residual effects from rebinning the energy derivatives (described in the text).  The broader peak in the correlation
function occurs at $\Delta t \sim$ -0.2 orbits.
\label{corr_z}}
\end{figure}

As discussed, the recurring channel modes in the net flux simulations
create distinguishable points of energy injection that make it
straightforward to follow the subsequent thermalization.  Such modes do
not exist in the zero net flux simulations, which makes the identification
of specific correlations slightly more difficult.  The situation is
further complicated by the overall reduced levels of the turbulence which
causes the time derivative of the thermal energy to be dominated by very high frequency oscillations due to
propagating spiral density waves \cite[][]{gard05b}.  We have determined
that these waves are created by compressibility and have very little effect
on the dissipational heating within the box.  
To remove their dominance in the energy derivatives,
 we rebin the time data using a ``neighborhood"
averaging procedure in which the rebinned data points are calculated from
averages of a specified number of original data points.  We then apply
equation~(\ref{corr_func}) between $\ein$ and $\td$; the result is shown
in Fig.~\ref{corr_z}.  The correlation curve has several narrow peaks,
which result from residual effects of the rebinning process.  The curve
has a broader peak near $\Delta t\sim$ -0.2~orbits, which agrees with the
same curves for $\none$ (Fig.~\ref{corr_n}).  The correlation
function for $\zone$ is not as sharply peaked as that for $\none$,
which is most likely a result of the lower amplitude variability in the
rebinned $\zone$ data.  Applying this analysis to the lower resolution
zero net flux simulations, we find that the correlation function always
has a broad peak at $\Delta t\sim$ -0.2~orbits.  Thus, as was the case in
the net flux simulations, the energy injection/thermalization timescale
is independent of resolution.

Finally, we note that the saturated state of $\zone$ is too complex to
obtain correlations between $\md$, $\kd$, and $\td$, such as was done
for $\none$.  In the net flux simulations, the recurring channel modes
lead to the build up and thermalization of magnetic energy.  The creation
and thermalization of magnetic energy are events that are well-separated
in time, making it easy to study the flow of energy between various
components.  In the zero net flux simulations, however, the average
properties of the turbulence remain more constant in time.  We will
further investigate the dissipation of magnetic and kinetic energy for the
zero net flux geometry in \S\ref{decaying} and \S\ref{zero_net_flux}.

\subsection{Decaying Turbulence}
\label{decaying}

As noted by \cite{haw95}, the MHD turbulence decays without differential
rotation to sustain the MRI.  We make use of this to observe how
rapidly thermalization occurs when there is no further input of energy.
This analysis should provide some additional
insight into the thermalization process for each field geometry.
We remove the net shear flow and the Coriolis and tidal forces from a
state taken from the sustained MRI turbulence in the fiducial models.
These runs are labeled ``NZD'' and ``SZD'' in Table~\ref{tbl:runs}  and are described
in more detail in \S\ref{method}.  Figure~\ref{turb_decay} shows
the subsequent magnetic and kinetic energy decay for both runs.
In the figure, the kinetic and magnetic energies have been normalized to
their values at the starting time of $t=40$ orbits.

In $\ndone$, the ratio of total magnetic to kinetic energy at $t=40$~orbits is 3.4.  The figure shows that
the magnetic energy decays more rapidly than the kinetic energy at early
times, losing almost half its initial value within 0.2 orbits.
In $\zdone$, the ratio of total magnetic to
kinetic energy at $t=40$~orbits is 1.4. The kinetic energy shows
high frequency oscillations about an average value that decays in time.
These oscillations are due to the same compressive, spiral waves that
exist in the sustained turbulence simulations.  The magnetic energy
is unaffected by these waves.  The average decay of kinetic energy,
calculated from smoothing away the oscillations, is also shown in the
figure.  Both the kinetic and magnetic energies decay quickly over time.
 Again, almost half the magnetic energy is lost within 0.2 orbits.
The high frequency oscillations also decay in amplitude over time. As
was the case in $\ndone$, the magnetic dissipation rate is initially
faster than that for the kinetic energy.  After about one orbit, the decay
rates become comparable.

Finally, we checked the contributions from the various terms in equations~(\ref{ke_eqn})
and (\ref{be_eqn}).  In both $\ndone$ and $\zdone$, there is some transfer from magnetic
to kinetic energy during the decay.  However, the transfer rate is small compared to the decay
rate of the magnetic energy and is such that the numerical dissipation of magnetic energy 
dominates over that of kinetic energy.

\section{Transfer Functions}
\label{trans_funcs}

In their investigation of convergence of zero net flux shearing box
simulations, \cite{from07a} carried out an analysis based on the evolution
of magnetic energy in Fourier space.  This analysis shows how magnetic
energy is created, transferred from one scale to another, and finally lost
due to numerical dissipation.  Their study used the ZEUS code and assumed
an isothermal equation of state.  Here we repeat and expand upon their
analysis to understand dissipation as a function of length scale in $\at$.

We note several differences between our work and that of \cite{from07a}.
First, they focus on magnetic energy evolution and did not provide a
comparable calculation for the kinetic energy.  Second, recognizing
that the $y$ direction is dominated by the largest scales, they restricted
their analysis to axisymmetric modes, namely $k_y = 0$.  Finally, as they
were primarily interested in how poloidal field could be regenerated as
part of a dynamo process, a portion of their analysis concentrated
on the poloidal components rather than the full magnetic field.  We have
chosen to extend the \cite{from07a} analysis more generally to include a
kinetic energy density evolution, nonaxisymmetric effects, and the effects 
of a nonzero toroidal field.

Following \cite{from07a}, we decompose the velocity field of the flow
into the mean flow, ${\bm V}_{\rm sh}$, and the turbulent velocity, $\vt$, via

\begin{equation}
\label{turb_velocity}
{\bm v} = {\bm V}_{\rm sh} + \vt.
\end{equation}

\noindent
The mean flow is defined as 

\begin{equation}
\label{vsh_define}
{\bm V}_{\rm sh} = \vsh \hat{\bm y} = \frac{\hat{\bm y}}{L_y L_z} \int \int v_y(x,y,z){\rm d}y
{\rm d}z.
\end{equation}

Turning next to the induction equation, we substitute
equation~(\ref{turb_velocity}) for the velocity, take the Fourier
transform, and dot the result with the complex conjugate of $\bfour$,
which is defined by equation~(\ref{four_trans}) with $f = {\bm B}$.
All Fourier transforms are done via equation~(\ref{four_trans}) using a
standard FFT and replacing $f$ with the appropriate quantity.  The data
is mapped into a frame in which the $x$ boundaries are periodic and then
remapped into the original frame after performing the FFT.

The result of this calculation is an equation describing the magnetic
energy density evolution in Fourier space,

\begin{equation}
\label{bfour_evolution}
\frac{1}{2}\frac{\partial|\bfour|^2}{\partial t} = A + S + \tbb + \tdivv + \tbv + \dmag,
\end{equation}

\noindent
where

\begin{equation}
\label{aterm}
A = -Re \left[\bstar \cdot \int \int \int \vsh \frac{\partial{\bm B}}{\partial y} \fterm \right],
\end{equation}

\begin{equation}
\label{sterm}
S = +Re \left[\widetilde{B}^\ast_{\it y}({\bm k}) \cdot \int \int \int B_x \frac{\partial \vsh}{\partial x} \fterm \right],
\end{equation}

\begin{equation}
\label{t_bb}
\tbb = -Re \left[\bstar \cdot \int \int \int (\vt \cdot \del){\bm B} \fterm \right],
\end{equation}

\begin{equation}
\label{t_divv}
\tdivv = -Re \left[\bstar \cdot \int \int \int (\del \cdot \vt){\bm B} \fterm \right],
\end{equation}

\begin{equation}
\label{t_bv}
\tbv = +Re \left[\bstar \cdot \int \int \int ({\bm B} \cdot \del)\vt \fterm \right].
\end{equation}

\noindent 
The $\dmag$ term has no analytic expression; it is simply
what is left over and accounts for numerical losses of magnetic energy
\cite[]{from07a}.  In the present simulations, there is no physical
resistivity.  The other terms have the following meanings: $A$ is the
transfer of magnetic energy between scales by the shear flow, $S$ is the
creation of magnetic energy from this shear flow, $\tbb$ is the advection of
magnetic energy between scales by the turbulent velocity field, $\tdivv$
results from the turbulent compressibility, and $\tbv$ describes the
creation of magnetic field by the turbulent velocity fluctuations.
In each case, $Re$ signifies the real part of the transform.

One can follow a similar procedure using the momentum equation to
determine the evolution of the kinetic energy density in Fourier space.
As described previously, we include the density in our Fourier transforms.
Consider the time derivative of $\sr{\bm v}$ given by

\begin{equation}
\label{rho_v_deriv1}
\frac{\partial\sr{\bm v}}{\partial t} = \sr \frac{\partial{\bm v}}{\partial t} + \frac{{\bm v}}{2\sr}\frac{\partial \rho}{\partial t}.
\end{equation}

\noindent 
Note that here, for simplicity, we do not decompose the velocity
into mean and turbulent components.  Using a combination of the continuity
and momentum equations, this equation can be written as

\begin{eqnarray}
\label{rho_v_deriv2}
\frac{\partial\sr{\bm v}}{\partial t} & = & \sr \left[-{\bm v} \cdot \del{\bm v} - \frac{1}{\rho}\del (P + \frac{1}{2}B^2) + \frac{1}{\rho}({\bm B} \cdot \del {\bm B}) - 2 {\bm \Omega} \times {\bm v} + 2 q \Omega^2 x \hat{{\bm x}} \right] \nonumber \\ 
& & + \frac{{\bm v}}{2 \sr} \left[-\rho(\del \cdot {\bm v}) - {\bm v} \cdot \del \rho \right],
\end{eqnarray}

\noindent 
If we take the Fourier transform of this equation and dot the
result with the complex conjugate of

\begin{equation}
\label{vfour}
\vfour = \int \int \int \sr({\bm x}) {\bm v}({\bm x}) \fterm,
\end{equation}

\noindent
we arrive at

\begin{equation}
\label{vfour_evolution}
\frac{1}{2}\frac{\partial|\vfour|^2}{\partial t} = \tvv + \tcomp + \tvb + \tpress + \tcor + \tphi + \dkin,
\end{equation} 

\noindent
where

\begin{equation}
\label{kin_fourier}
\frac{1}{2}|\vfour|^2 \equiv \frac{1}{2}\left[|\widetilde{\sqrt{\rho}v_x({\bm k})}|^2+|\widetilde{\sqrt{\rho}v_y({\bm k})}|^2+|\widetilde{\sqrt{\rho}v_z}({\bm k})|^2\right],
\end{equation} 

\begin{equation}
\label{t_vv}
\tvv = -Re \left[\vstar \cdot \int \int \int [\sr({\bm v} \cdot \del){\bm v} + \frac{{\bm v}}{2 \sr}({\bm v} \cdot \del)\rho] \fterm \right],
\end{equation}

\begin{equation}
\label{t_comp}
\tcomp = -Re \left[\vstar \cdot \int \int \int \frac{\sr{\bm v}}{2} (\del \cdot {\bm v}) \fterm \right],
\end{equation}

\begin{equation}
\label{t_vb}
\tvb = +Re \left[\vstar \cdot \int \int \int \frac{1}{\sr}({\bm B} \cdot \del) {\bm B} \fterm \right],
\end{equation}

\begin{equation}
\label{t_press}
\tpress = -Re \left[\vstar \cdot \int \int \int \frac{1}{\sr}\del(P + \frac{1}{2}B^2) \fterm \right],
\end{equation}

\begin{equation}
\label{t_cor}
\tcor = -Re \left[\vstar \cdot \int \int \int (2{\bm \Omega} \times \sr {\bm v}) \fterm \right],
\end{equation}

\begin{equation}
\label{t_phi}
\tphi = +Re \left[\widetilde{\sqrt{\rho}v_x^{\bm \ast}}({\bm k}) \cdot \int \int \int 2 \sr q \Omega^2 x \fterm \right],
\end{equation}

\noindent
and $\dkin$ accounts for the dissipation of kinetic energy. Again,
this dissipation is numerical as we have not included an explicit
viscosity term in our equations. Equation~(\ref{vfour_evolution})
describes the evolution of the kinetic energy density in Fourier space.
$\tvv$ is a term that describes the transfer of kinetic energy between
scales by the velocity field (both the mean and turbulent velocity),
$\tcomp$ results from turbulent compressibility, $\tvb$ describes how
kinetic energy changes from magnetic tension, $\tpress$ represents the
effect of both gas and magnetic pressure on the kinetic energy, and $\tphi$
is the effect of the tidal potential on the kinetic energy.  Note that $\tcor$
is analytically equal to zero, and it is not included in any of
the following analysis or discussion.

In the saturated state of the MRI, the magnetic and kinetic energy
densities should be in a steady state on average (although they
do show strong fluctuations over short periods of time).   If we
consider the time-averages of equations~(\ref{bfour_evolution}) and
(\ref{vfour_evolution}), then we can set the left hand sides to zero.
We then rewrite these equations as

\begin{equation}
\label{vfour_zero}
\tvv + \tcomp + \tvb + \tpress + \tphi + \dkin = 0,
\end{equation}

\begin{equation}
\label{bfour_zero}
A + S + \tbb + \tdivv + \tbv + \dmag = 0,
\end{equation}

\noindent
where each of these terms is now a time-average.  Here we average 
over 161 snapshots from orbit 20 to 100 in increments of 0.5 orbits.
Each of these terms is a function of $k_x, k_y,$ and $k_z$,
and in what follows we average the terms on shells of constant $k = |{\bm
k}|$ as was done in \cite{from07a}.\footnote{In our analysis, the average over shells of
constant $k$ was done before the temporal average.}  Note that unlike the averaging
described in that paper, we include $k_y$ in the calculation of $k$.

\subsection{Zero Net Magnetic Flux}
\label{zero_net_flux}

\subsubsection{Fiducial Run}
\label{fid_z}

\begin{figure}
\plotfiddle{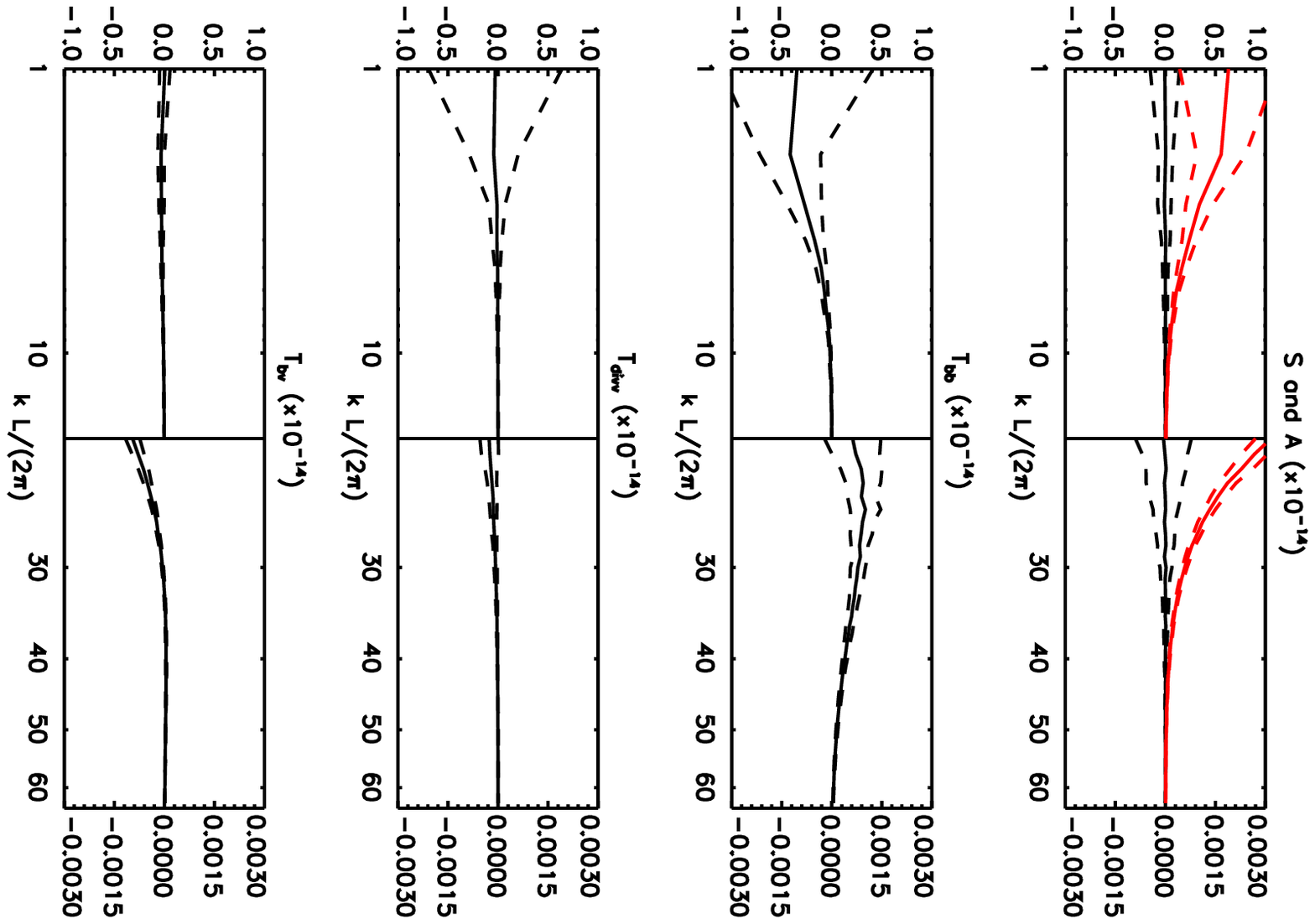}{0.1in}{90.}{475}{300}{80}{-500}
\caption{Magnetic Fourier transfer functions versus a dimensionless wave number ($L$ is the length of the smallest dimension in the box) for $\zone$. Each plot
is displayed in two components; the left part shows the data for $1 < k L/(2\pi) < 20$, and the right part shows the data for $20 < k L/(2\pi) < 64$ by
changing the $x$ and $y$ axis scaling.  In all plots, the solid line is the average value for the transfer function.  This average was obtained over 161 frames in the saturated state and shells of constant $|{\bm k}|$.  The upper (lower) dashed line that matches color with the solid line correspond to the transfer function plus (minus) one temporal standard deviation.  From top to bottom, the plots show $S$ (red) and $A$ (black), $\tbb$, $\tdivv$, and $\tbv$.
\label{magtfz}}
\end{figure}

\begin{figure}
\plotfiddle{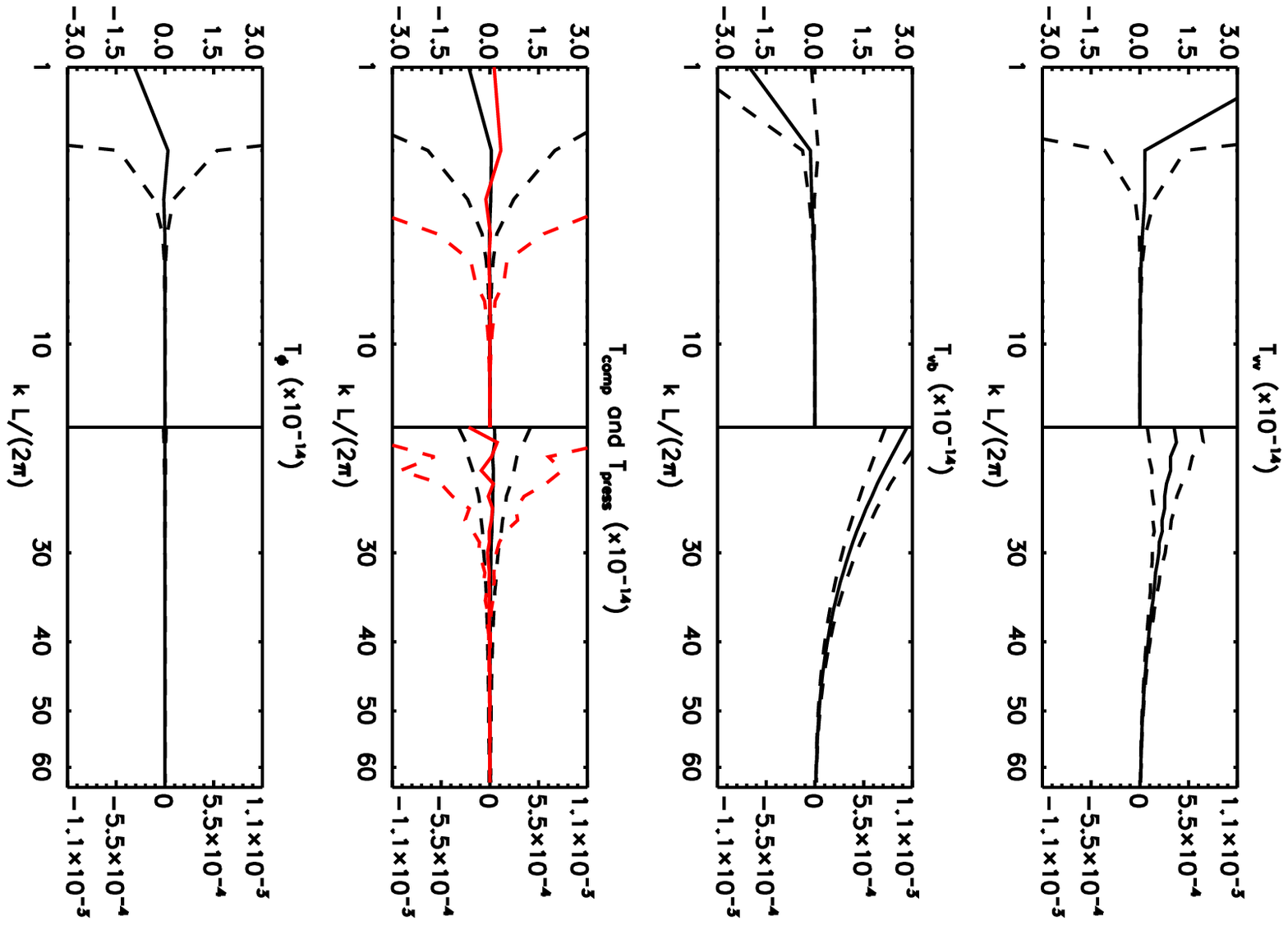}{0.1in}{90.}{475}{300}{80}{-500}
\caption{Kinetic Fourier transfer functions versus a dimensionless wave number ($L$ is the length of the smallest dimension in the box) for $\zone$. Each plot
is displayed in two components; the left part shows the data for $1 < k L/(2\pi) < 20$, and the right part shows the data for $20 < k L/(2\pi) < 64$ by
changing the $x$ and $y$ axis scaling.  In all plots, the solid line is the average value for the transfer function.  This average was obtained over 161 frames in the saturated state and shells of constant $|{\bm k}|$.  The upper (lower) dashed line that matches color with the solid line correspond to the transfer function plus (minus) one temporal standard deviation.  From top to bottom, the plots show $\tvv$, $\tvb$, $\tpress$ (red) and $\tcomp$ (black), and $\tphi$.
\label{kintfz}}
\end{figure}

In this section, we focus on the Fourier transfer
functions for the fiducial zero net magnetic flux simulation.
Figure~\ref{magtfz} plots the magnetic transfer functions
defined in equations~(\ref{aterm})-(\ref{t_bv}) as a function of length
scale for $\zone$, and Fig.~\ref{kintfz} plots the kinetic
transfer functions defined in equations~(\ref{t_vv})-(\ref{t_phi}).
The dashed lines correspond to plus or minus one standard deviation around
the mean value of the time average.  Most of the transfer functions show
large variation at small $k$ values which may be due to poor statistics
at small $k$ and relatively large time variability.  Because the transfer
functions approach zero rapidly, we plot the ranges $1 < k L/(2\pi) < 20$ 
and $20 < k L/(2\pi) < 64$ in the same figure, but with different $y$ scalings.

The shear term $S$ is positive at all scales, as observed in
\cite{from07a}, meaning that $B_y$ is created by the shear flow at
all scales.  $A$ is small at all scales, supporting the assumption made
in \cite{from07a} that $A \approx 0$.  $\tbv$ is primarily negative
at the largest scales, although there are large fluctuations, and
becomes positive for $k L/(2\pi) \gtrsim 35$.  The turbulent velocity
fluctuations seem to be creating magnetic energy at the smallest scales,
but at larger scales, the magnetic field appears to lose energy via this
interaction with the turbulence.    $\tbb$ is negative for $k$ smaller
than $k L/(2\pi) \sim 20$, meaning that the turbulence is transferring
magnetic energy away from these scales.  Although this analysis doesn't
determine the direction of this cascade, at the largest scale (i.e.,
the box size) the energy
can only cascade to smaller scales.  In terms of absolute value, $S$
and $\tbb$ are dominant on the largest scales, while on small scales,
$\tbb > \tbv > S > 0$.

It is difficult to say anything conclusive about the kinetic transfer
functions on the largest scales as they are subject to considerable
uncertainty, although $\tvb < 0$ appears reasonably well constrained
at these scales. At smaller scales, the two dominant terms are $\tvv$
and $\tvb$, with $\tvb > \tvv > 0$; kinetic energy is being transferred
to these scales by the turbulence, and being created by magnetic field.

\begin{figure}
\plotfiddle{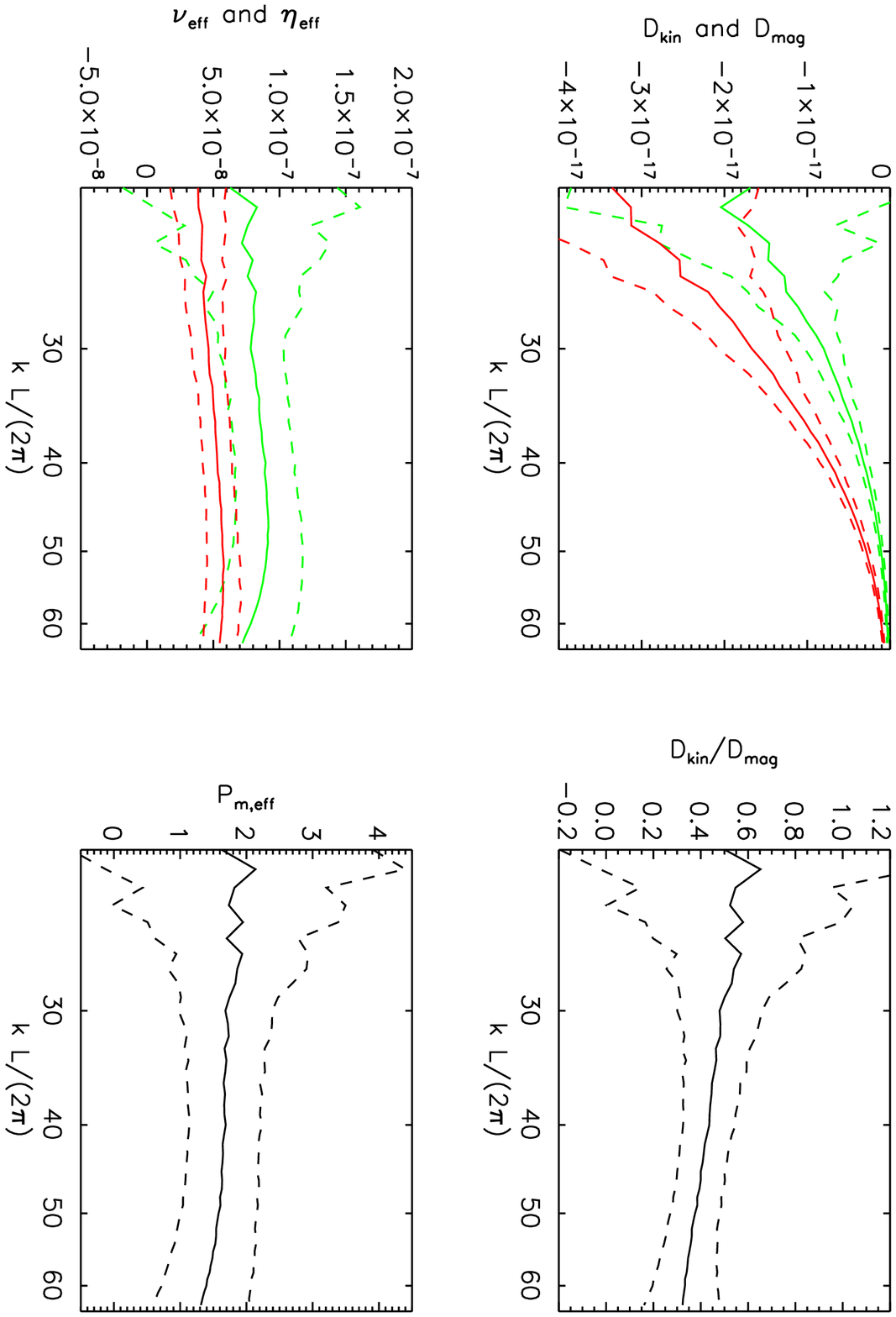}{1in}{90.}{325}{475}{-10}{-100}
\vspace{-1in}
\caption{Numerical dissipation quantities plotted against a dimensionless wave number ($L$ is the length of the smallest dimension in the box). These plots correspond to data from $\zone$.  The upper left plot shows the dissipation rate of kinetic energy (green) and magnetic energy (red) in Fourier space.  The upper right plot shows the ratio of these two dissipation rates.  The lower left plot shows the effective numerical viscosity (green) and resistivity (red).  The lower right plot shows the ratio of the viscosity to resistivity (i.e., the effective Prandtl number).   In all plots, the solid line is the average value for the quantity of interest.  For $\dkin$ and $\dmag$, this average was obtained from averaging over shells of constant $|{\bm k}|$ and over 161 frames in the saturated state. The averaged viscosity and resistivity values were calculated as described in the text. The upper and lower dashed lines correspond to the error propagated from one temporal standard deviation.
\label{dis_z}}
\end{figure}

Equations~(\ref{vfour_zero})~and~(\ref{bfour_zero}) have been set to
zero from the assumption that the magnetic and kinetic energies are in a
time-averaged steady state.  The dissipation terms $\dmag$ and $\dkin$
are simply what is left over after the other transfer functions have
been computed.  The top plots in Fig.~\ref{dis_z} are the kinetic and
magnetic dissipation and the ratio $\dkin/\dmag$ as a function of $k$
for $20 < k L/(2\pi) < 64$; the scatter at small $k$ is large and there
is considerable uncertainty in the dissipation values.  At small scales,
magnetic dissipation dominates kinetic dissipation by a factor of roughly
three. The kinetic and magnetic dissipation rate increases in magnitude
towards larger scales.

Following \cite{from07a}, we can determine an effective resistivity and
viscosity as a function of length scale by assuming that the numerical
effects behave as if they were physical resistivity and viscosity.
For example, with a constant Ohmic resistivity, the induction equation
would have an additional term proportional to $\nabla^2 B$, with the
constant of proportionality being the resistivity.  If we take the Fourier
transform of this term and dot it with the complex conjugate of $\bfour$,
the real part is

\begin{equation}
\label{t_eta}
\teta = +Re \left[\bstar \cdot \int \int \int \nabla^2{\bm B} \fterm \right] = -k^2|\bfour|^2.
\end{equation}

\noindent
We can then define an effective resistivity as a function of $k$ by

\begin{equation}
\label{eta_eff}
\etaeff(k) \equiv \frac{\dmag(k)}{\teta(k)}.
\end{equation}

Similarly, a constant kinematic shear viscosity would add a term
proportional to \\ $\sr [\nabla^2{\bm v} + \frac{1}{3}\del(\del \cdot {\bm v})]$ to equation~(\ref{rho_v_deriv2}), with the constant of
proportionality being the viscosity.  Note that we only consider shear
viscosity here for simplicity.  We take the Fourier transform of the
viscous term, dot it with the complex conjugate of equation~(\ref{vfour}),
and take the real part.  The result is

\begin{equation}
\label{t_nu_first}
\tnu = +Re \left[\vstar \cdot \int \int \int \sr [\nabla^2{\bm v} + \frac{1}{3}\del(\del \cdot {\bm v})] \fterm \right].
\end{equation}

\noindent 
This equation can be made simpler by realizing that the
second term of the integrand, related to the divergence of ${\bm v}$,
is negligible.  We can also assume that the density is relatively
constant, and arrive at

\begin{equation}
\label{t_nu}
\tnu = -k^2|\vpfour|^2 .
\end{equation}

\noindent 
We have substituted the perturbed velocity here because it
is the only velocity that can lead to {\it numerical} dissipation of
kinetic energy.  That is, a pure shear flow will not encounter any numerical
viscosity, and we can subtract off this flow.  We define an effective
viscosity by

\begin{equation}
\label{nu_eff}
\nueff(k) \equiv \frac{\dkin(k)}{\tnu(k)}.
\end{equation}

We can also characterize the effective
resistivity and viscosity in terms of a Reynolds number,

\begin{equation}
\label{re}
\reeff(k) \equiv \frac{c_o H}{\nueff(k)}, 
\end{equation}

\noindent
and magnetic Reynolds number,

\begin{equation}
\label{rm}
\rmeff(k) \equiv \frac{c_o H}{\etaeff(k)}, 
\end{equation}

\noindent
where we have used the initial isothermal sound speed, $c_o = 0.001$, as
a characteristic velocity, and $H = L_z$ is a characteristic length. 
These numbers quantify the numerical dissipation coefficients
in a dimensionless manner.

Finally, we define an effective Prandtl number by

\begin{equation}
\label{pmeff}
\pmeff(k) \equiv \frac{\nueff(k)}{\etaeff(k)}
\end{equation}

The effective viscosity and resistivity as well as the effective Prandtl
number are shown in the bottom plots of Fig.~\ref{dis_z}.  
The viscosity and resistivity are fairly constant at large $k$.
The effective Reynolds numbers are on the order of $\reeff \sim 12000$, and $\rmeff \sim 20000$ at large $k$.
The Prandtl number is also relatively flat at these scales, and $\pmeff
\sim 1.6$.  This result agrees with \cite{from07a}, where $\pmeff >
1$ for ZEUS. While the numerical dissipation of $\at$ is not physical,
the ``flatness" of $\nueff$ and $\etaeff$ suggests a resemblance to
physical dissipation at small scales.  

Finally, note that although the Prandtl number is greater than unity,
the magnetic dissipation dominates over kinetic dissipation.  Evidently,
$\teta$ is larger than $\tnu$ because there is more magnetic energy than
kinetic energy at a given scale.  In particular,

\begin{equation}
\label{energy_ratio1}
\frac{\teta}{\tnu} = \frac{|\bfour|^2}{|\vpfour|^2}.
\end{equation}

\noindent
Since there is more magnetic energy than perturbed kinetic energy at a 
given scale, magnetic dissipation dominates.

\subsubsection{Resolution Effects}
\label{res_effects_z}

\begin{figure}
\plotfiddle{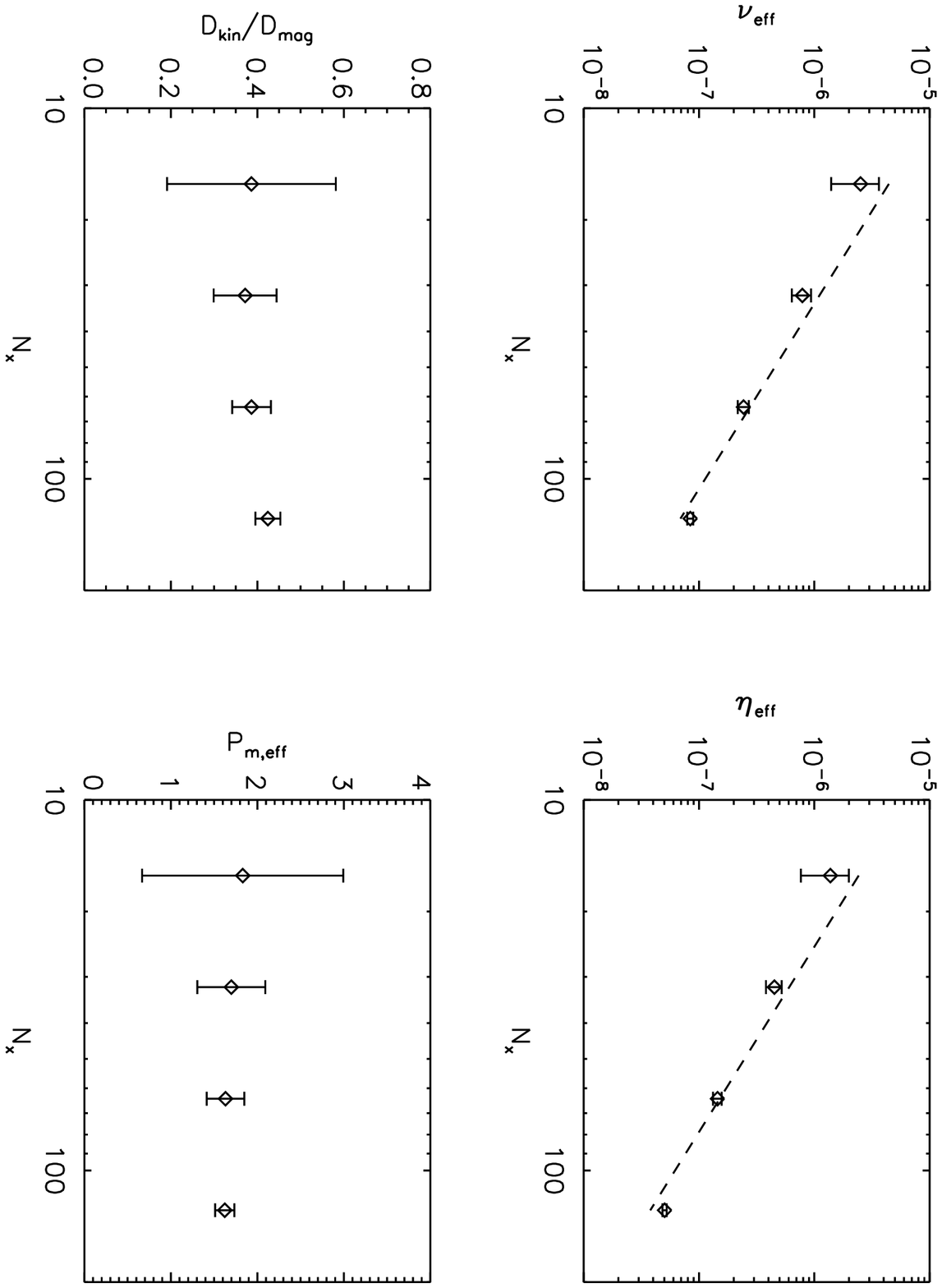}{1in}{90.}{325}{475}{-10}{-100}
\vspace{-1in}
\caption{Averaged dissipation related quantities as a function of grid resolution.  These plots correspond to data from the zero net flux simulations, $\zonesix$, $\zthree$, $\zsix$, and $\zone$.   The upper left plot shows the effective viscosity versus $x$ resolution. The dashed line shows $\nueff \propto N_x^{-2}$.  The upper right plot shows the effective resistivity versus $x$ resolution. Again, the dashed line shows $\etaeff \propto N_x^{-2}$. The lower left plot shows the ratio of kinetic to magnetic dissipation versus $x$ resolution.  The lower right plot shows the effective Prandtl number versus $x$ resolution. For each resolution, the data point was obtained from averaging the quantity as a function of $k$ over values of $k$ where the error in this quantity is not much larger than the mean value. The error bars represent the propagated errors from
the temporal statistics.
\label{nu_eta_res_z}}
\end{figure}

To gauge the effect of resolution on these various quantities, we
perform the same analysis on the lower resolution runs, $\zonesix$,
$\zthree$, and $\zsix$. We focus, in particular, on the small scales 
(i.e., large $k$) where our quantities are statistically more well-determined.
Figure~\ref{nu_eta_res_z} shows $\nueff$, $\etaeff$, $\pmeff$, and the
ratio of $\dkin$ to $\dmag$ as a function of $x$ resolution, $N_x$. The
data points are calculated by averaging the quantity of interest over $k$
in the regions of $k$-space where the error on the quantity is less than
its mean value.\footnote{There are some quantities for which the error is
never less than the mean.  In these cases, we average over regions where
the mean is greater than 80\% of the error.} The displayed error bars are
the propagation of the errors from the temporal statistics.  At these
large values of $k$, $\nueff$, $\etaeff$, $\pmeff$, and $\dkin/\dmag$
are relatively flat, varying by a factor of at most 2. Consequently,
these averages should be representative at small scales.

The numerical viscosity and resistivity decrease as a function of
resolution. The dashed lines in the two upper panels of the figure
show the line $\nueff, \etaeff \propto N_x^{-2}$.  The viscosity and
resistivity decrease slower than this with increasing $N_x$; we measured
$\nueff, \etaeff \propto N_x^{-1.6}$. The figure also shows that both the
effective Prandtl number and the ratio of kinetic to magnetic dissipation
are constant with resolution to within the error bars.

\subsubsection{Comparison with Previous Results}
\label{trans_compare_section}

\cite{from07a} were interested in the transfer function for the poloidal
field, as the regeneration of this field is key to a self-sustaining
dynamo.  They found that the magnetic dissipation of ZEUS for the
poloidal magnetic field departs from the physical dissipation model
at small $k$ and could even be a nonphysical ``positive" dissipation.
We repeat the same analysis as performed in that paper, but with $\zone$,
for comparison.  First, we examine the magnetic dissipation for the full
3D Fourier analysis described above.  Second, we do the same procedure but
setting $B_y = 0$ to focus on the effect of only including poloidal field.
Finally, we perform the procedure with $B_y = 0$ and in the plane $k_y
= 0$ (i.e., axisymmetry).  These simplifications allow us to reproduce
the poloidal field analysis of \cite{from07a}.

\begin{figure}
\plotfiddle{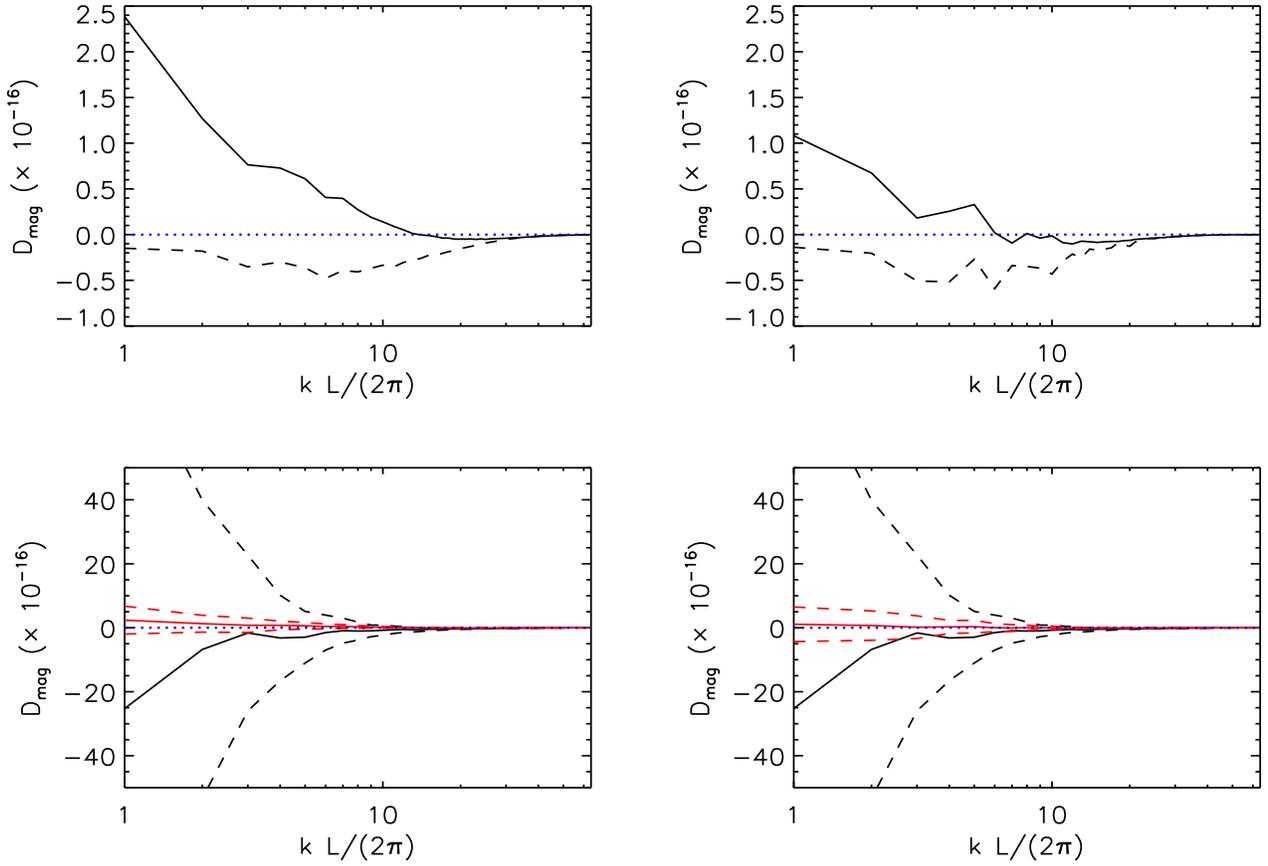}{1in}{90.}{325}{475}{-10}{-100}
\vspace{-1in}
\caption{Magnetic dissipation rate from the $\zone$ simulation for three versions of the transfer function analysis.  The upper left plot and the red lines in the lower left plot correspond to the analysis in which $B_y = 0$ was assumed.  The upper right plot and the red lines in the lower right plot correspond to the analysis in which both $B_y = 0$ and $k_y = 0$ were assumed.  The black lines in the lower plots result from relaxing both of these assumptions.  The solid lines in the upper plots correspond to $\dmag$ whereas the dashed lines correspond to $\eta \teta$ with $\eta = 10^{-7}$ chosen to provide a reasonable match to $\dmag$ at large $k$.  The dashed lines in the lower plots correspond to one standard deviation above and below the quantity represented by the solid line of the same color.  A horizontal line at zero is shown in all plots as the blue dotted line.  Note the difference in $y$-axis scale between the upper and lower plots.
\label{trans_compare}}
\end{figure}

The results are shown in Fig.~\ref{trans_compare}.  The left two plots
correspond to the Fourier analysis in which only $B_y = 0$ is assumed.
The right plots assume $B_y = 0$ and $k_y = 0$. The black lines in the
bottom two plots correspond to the magnetic dissipation for the full 3D
Fourier analysis with $B_y \neq 0$ and $k_y \neq 0$. It is apparent that
when $B_y = 0$ is assumed in the calculations, the magnetic dissipation
becomes positive at large scales.  However, when $B_y$ is included, the
magnetic dissipation remains negative.  Whether or not $k_y = 0$ is
assumed seems to make very little difference, supporting the notion that
small $k_y$ dominates.  Since $\at$ and ZEUS both find positive $\dmag$ at
small $k$, it is unlikely that this effect can be attributed to algorithmic
limitations specific to ZEUS.  Since $\dmag$ is not a derived quantity but
simply what remains after all the transfer functions are calculated, 
it seems likely that the positive $\dmag$ values for the poloidal field analysis
are due to incomplete statistics at large scales, or other inadequacies of the 
analysis when applied solely
to the poloidal field.  At small $k$, the standard deviations of the
quantities (dashed lines) are considerable.  The standard deviation on
$\dmag$ when $B_y \neq 0$ is significantly larger than when one sets
$B_y = 0$.   This reflects the large variability of $\langle B_y^2/2
\rangle$ compared to the other components of magnetic energy (see e.g.,
Fig.~\ref{turb_z}).  At any given time, $\dmag$ can be positive;
the assumption of time-stationarity does not hold at any point in time.
But when the data are time-averaged, $\dmag < 0$.

Finally, we compare the numerical magnetic Reyolds number calculated with
equation~(\ref{rm}) but with the $B_y = 0$ and $k_y = 0$ assumptions.
For $\zone$, we find that $\rmeff \sim$ 11000, and for $\zsix$, $\rmeff
\sim$ 3500.  \cite{from07a} find $\rmeff \sim$ 30000 for their $N_x =
128$ run, and $\rmeff \sim$ 10000 for their $N_x = 64$ run; both of their
calculated effective Reynolds numbers are larger than those calculated
for $\at$.  This result seems to suggest that ZEUS is actually less
dissipative than $\at$.  However, there are several points to consider.
First, numerical dissipation is a nonlinear function of resolution,
sharply increasing as the number of zones per wavelength decreases
(high wavenumbers).  The effective Reynolds number is obtained by
measuring dissipation at the high $k$ end of the spectrum.  As reported
by \cite{shen06} $\at$ appears to have higher dissipation than ZEUS
for poorly resolved waves, as evidenced by the ability of $\at$ to
avoid the aliasing errors seen with ZEUS for hydrodynamic shearing
box waves.  They further point out that for wavelengths larger than
16 grid points $\at$ is less dissipative.  Further, 2D simulations of
decaying turbulence have demonstrated that when saturation amplitude is reached,
the decay time is longer in $\at$ than in ZEUS, consistent with
$\at$ having a higher effective resolution \cite[]{stone05}.
In the present context, we find that the time- and volume-averaged total
stresses in our simulations are larger than those calculated in the
simulations of \cite{from07a}.  Stronger turbulence leads to larger kinetic
and magnetic turbulent fluctuations, which in turn enhances dissipation
via grid-scale effects.  Finally, we reemphasize that assuming $B_y = 0$
may have a significant impact on the measurement of effective magnetic
dissipation via this analysis.

\subsection{Net Magnetic Flux}
\label{net_flux}

\subsubsection{Fiducial Run}
\label{fid_n}

\begin{figure}
\plotfiddle{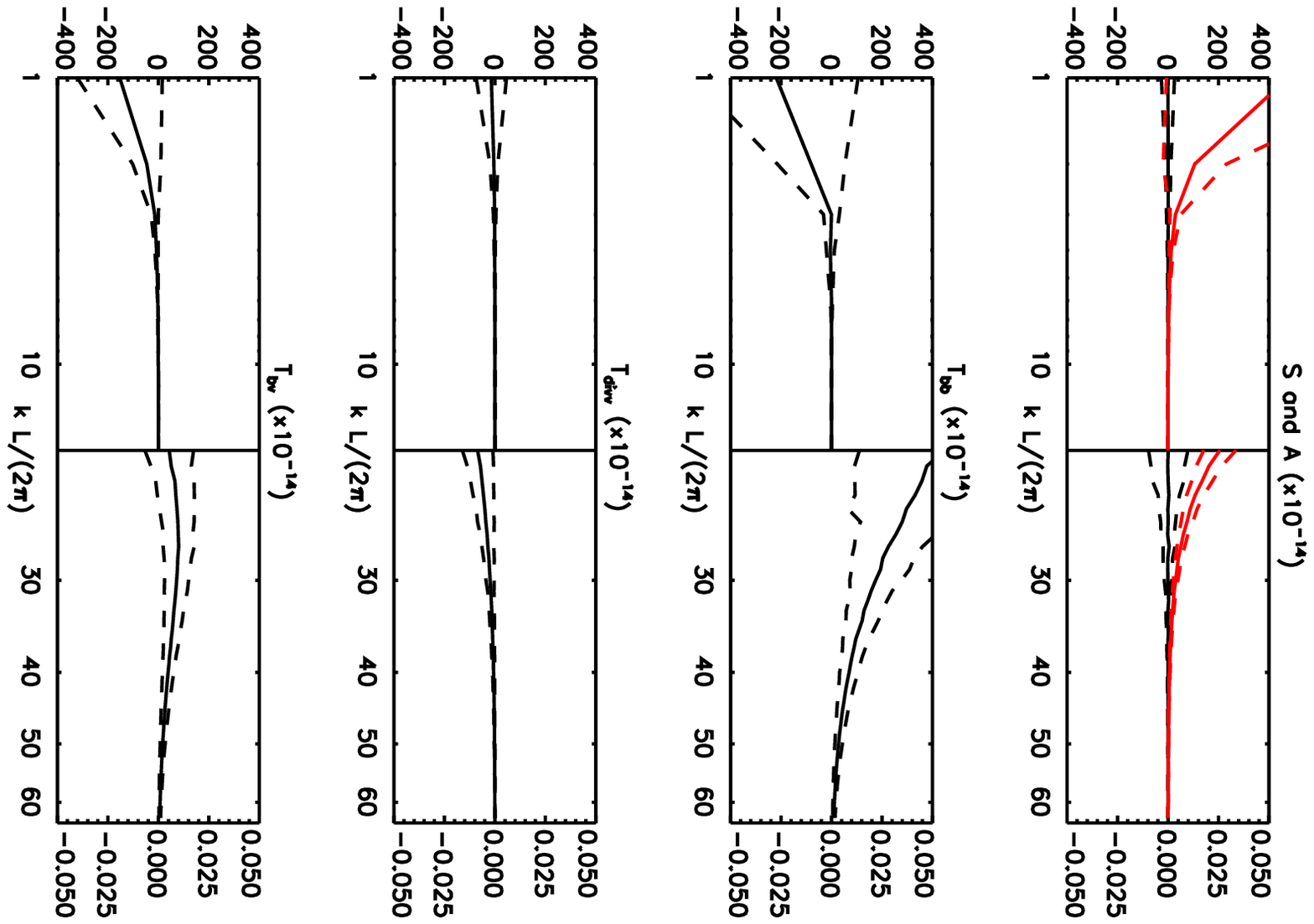}{0.1in}{90.}{475}{300}{80}{-500}
\caption{Magnetic Fourier transfer functions versus a dimensionless wave number ($L$ is the length of the smallest dimension in the box) for $\none$. Each plot
is displayed in two components; the left part shows the data for $1 < k L/(2\pi) < 20$, and the right part shows the data for $20 < k L/(2\pi) < 64$ by
changing the $x$ and $y$ axis scaling.  In all plots, the solid line is the average value for the transfer function.  This average was obtained over 161 frames in the saturated state and shells of constant $|{\bm k}|$.  The upper (lower) dashed line that matches color with the solid line correspond to the transfer function plus (minus) one temporal standard deviation.  From top to bottom, the plots show $S$ (red) and $A$ (black), $\tbb$, $\tdivv$, and $\tbv$.
\label{magtfn}}
\end{figure}

\begin{figure}
\plotfiddle{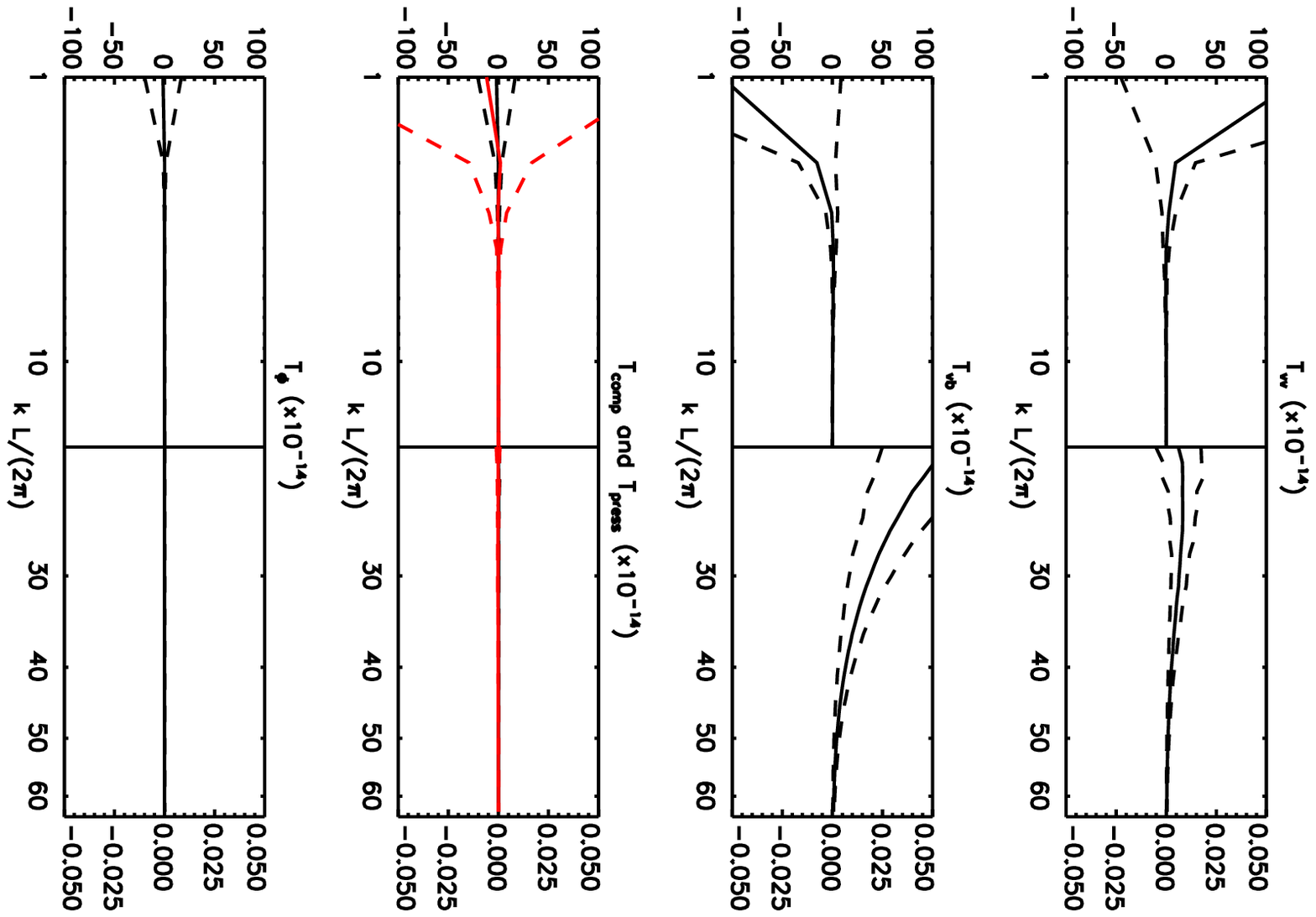}{0.1in}{90.}{475}{300}{80}{-500}
\caption{Kinetic Fourier transfer functions versus a dimensionless wave number ($L$ is the length of the smallest dimension in the box) for $\none$. Each plot
is displayed in two components; the left part shows the data for $1 < k L/(2\pi) < 20$, and the right part shows the data for $20 < k L/(2\pi) < 64$ by
changing the $x$ and $y$ axis scaling.  In all plots, the solid line is the average value for the transfer function.  This average was obtained over 161 frames in the saturated state and shells of constant $|{\bm k}|$.  The upper (lower) dashed line that matches color with the solid line correspond to the transfer function plus (minus) one temporal standard deviation.  From top to bottom, the plots show $\tvv$, $\tvb$, $\tpress$ (red) and $\tcomp$ (black), and $\tphi$.
\label{kintfn}}
\end{figure}

We perform the same transfer function analysis on the
fiducial net magnetic flux run, $\none$. The various
transfer function terms as a function of $k$ are shown in
Figs.~\ref{magtfn}--\ref{kintfn}. As was the
case in the zero net flux simulation, $S$ is positive at all scales
and dominates at small $k$; $A$ is relatively small throughout. $\tbv$
and $\tbb$ are negative at large scales and positive at small scales,
with $\tbb > 0$ for $k L/(2\pi) \gtrsim 5$, and $\tbv >0$ for $k L/(2\pi)
\gtrsim 20$.   At small scales, $\tbb > \tbv > S > 0$.  Of the kinetic
terms, $\tvv$ and $\tvb$ dominate with $\tvb > \tvv > 0$.  These results
are in general agreement with $\zone$, except that the magnitude of
the various terms is larger for $\none$ than for $\zone$, and $\tbb$
and $\tbv$ become positive at smaller $k$ values compared to $\zone$.

\begin{figure}
\plotfiddle{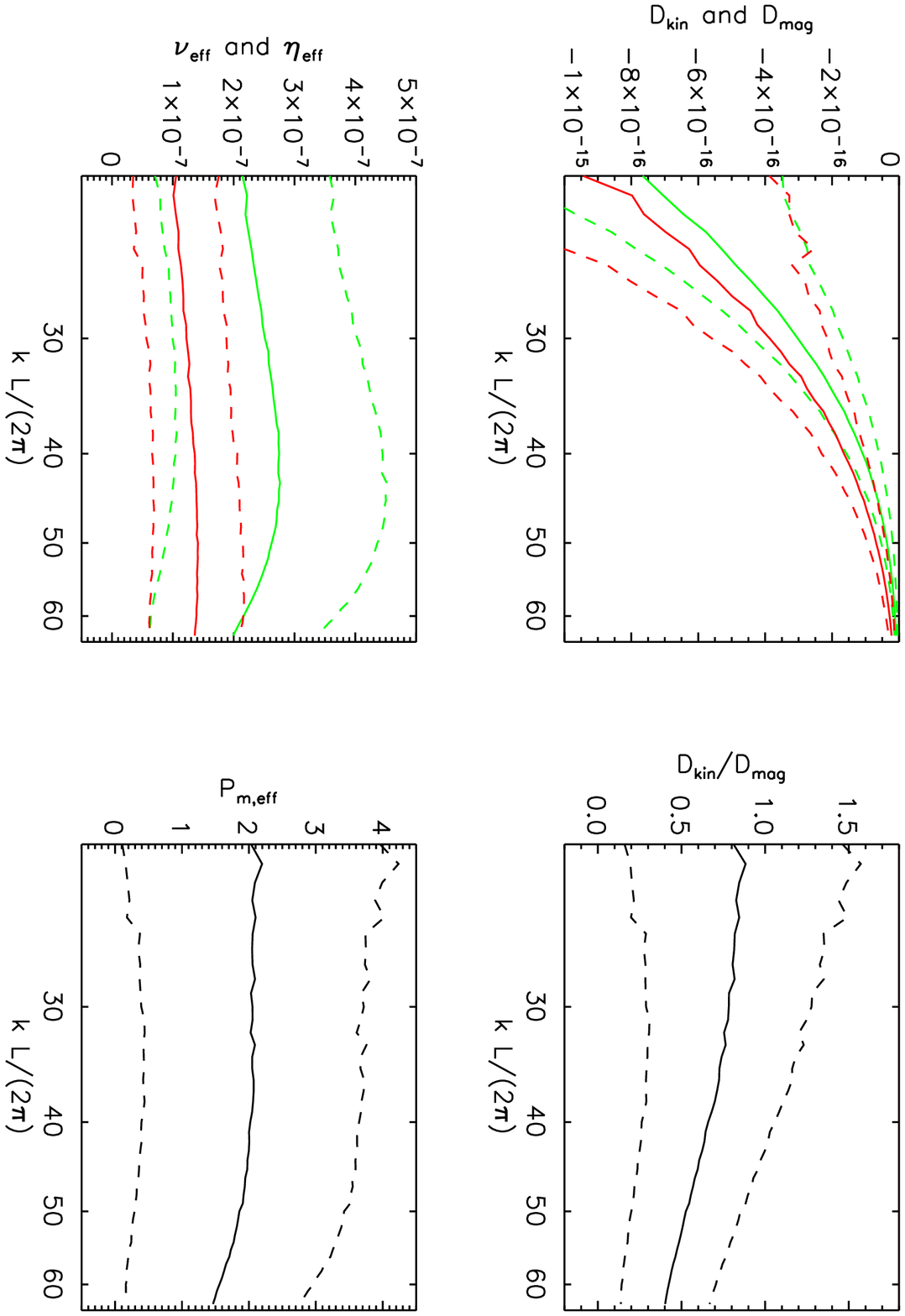}{1in}{90.}{325}{475}{-10}{-100}
\vspace{-1in}
\caption{Numerical dissipation quantities plotted against a dimensionless wave number ($L$ is the length of the smallest dimension in the box). These plots correspond to data from $\none$.  The upper left plot shows the dissipation rate of kinetic energy (green) and magnetic energy (red) in Fourier space.  The upper right plot shows the ratio of these two dissipation rates.  The lower left plot shows the effective numerical viscosity (green) and resistivity (red).  The lower right plot shows the ratio of the viscosity to resistivity (i.e., the effective Prandtl number).   In all plots, the solid line is the average value for the quantity of interest.  For $\dkin$ and $\dmag$, this average was obtained from averaging over shells of constant $|{\bm k}|$ and over 161 frames in the saturated state. The averaged viscosity and resistivity values were calculated as described in the text. The upper and lower dashed lines correspond to the error propagated from one temporal standard deviation. 
\label{dis_n}}
\end{figure}

As before, we calculate the kinetic and magnetic dissipation as well as
effective values for the viscosity and resistivity.  Figure~\ref{dis_n}
shows these quantities for $\none$ at the smallest scales.   As was the
case for $\zone$, the mean magnetic dissipation dominates over kinetic
dissipation by a factor of roughly three at these scales.  Note, however,
the large error bars associated with these plots, which encompass values
of $\dkin/\dmag > 1$.  Again, the error bars are the temporal 
standard deviation of the transfer functions.
Since $\none$ has a larger temporal variability, larger error bars
are expected.  The mean value for $\dkin/\dmag$ is on the order of
0.6-0.7, which is consistent with the analysis in \S~\ref{sustained} in which
we found $\qk/\qm \sim 0.6$.

The effective viscosity and resistivity show the same basic result as
in the $\zone$ case.  $\nueff$, $\etaeff$, and $\pmeff$ change by a
factor of order unity at large $k$.  The effective
Reynolds numbers are on the order of $\reeff \sim 4000$, and $\rmeff \sim 8000$ at large $k$.
$\pmeff$ has a mean value of $\sim 1.9$.  Again, there is considerable uncertainty in these values
due to the large amplitude fluctuations in the turbulence.  The error
bars encompass values of $\pmeff$ less than unity.  As a result, it
is more difficult to conclusively say that the dissipation behaves the
same way in $\none$ as in $\zone$.  However, in an average sense, the
two simulations agree well qualitatively.

\subsubsection{Resolution Effects}
\label{res_effects_n}

\begin{figure}
\plotfiddle{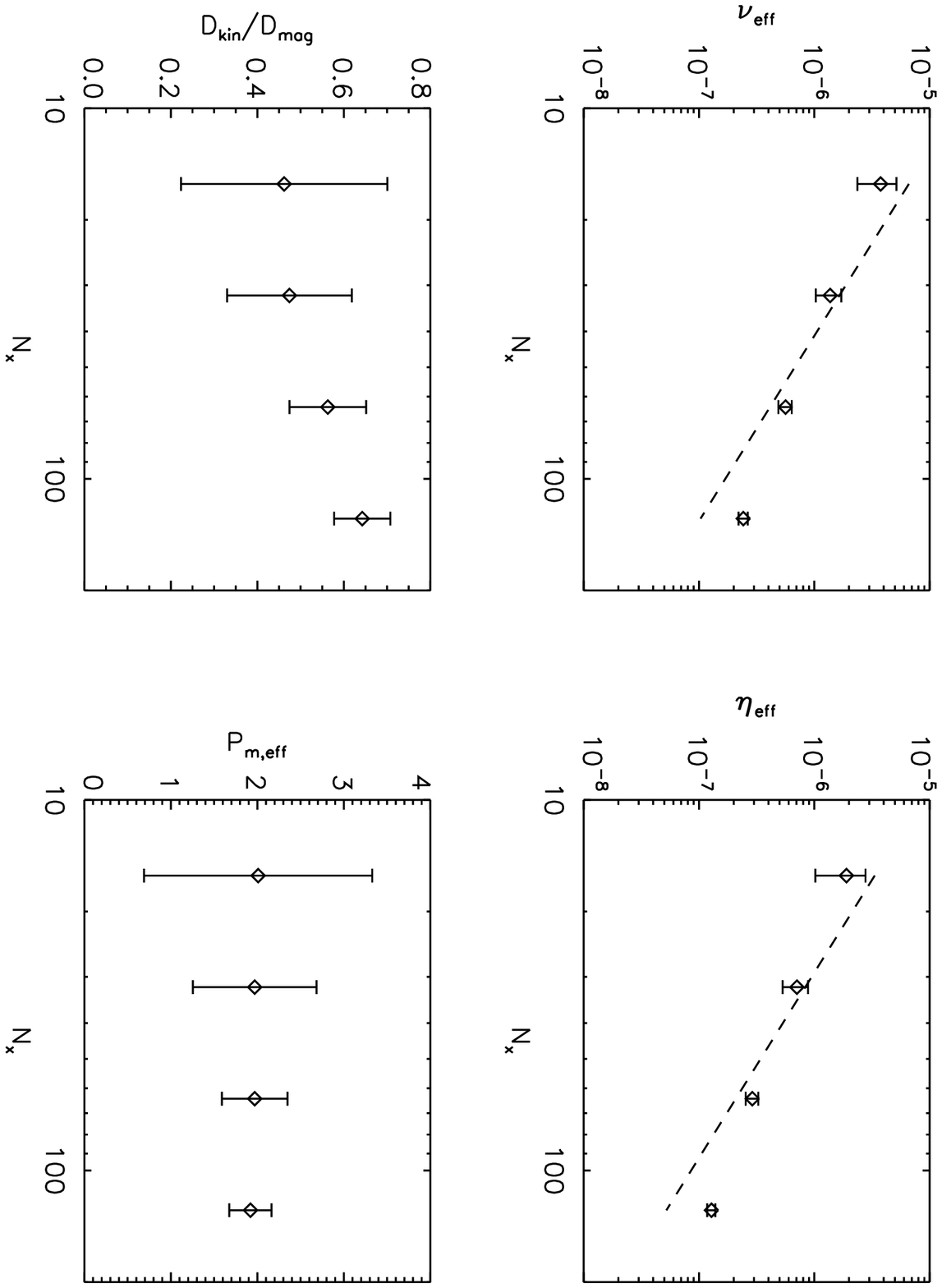}{1in}{90.}{325}{475}{-10}{-100}
\vspace{-1in}
\caption{Averaged dissipation related quantities as a function of grid resolution.  These plots correspond to data from the net flux simulations, $\nonesix$, $\nthree$, $\nsix$, and $\none$.   The upper left plot shows the effective viscosity versus $x$ resolution. The dashed line shows $\nueff \propto N_x^{-2}$.  The upper right plot shows the effective resistivity versus $x$ resolution. Again, the dashed line shows $\etaeff \propto N_x^{-2}$. The lower left plot shows the ratio of kinetic to magnetic dissipation versus $x$ resolution.  The lower right plot shows the effective Prandtl number versus $x$ resolution. For each resolution, the data point was obtained from averaging the quantity as a function of $k$ over values of $k$ where the error in this quantity is not much larger than the mean value. The error bars represent the propagated errors from
the temporal statistics.
\label{nu_eta_res_n}}
\end{figure}

We can again look at the effect of resolution on these various dissipation
quantities.  Figure~\ref{nu_eta_res_n} shows this effect for the net flux
simulations ($\nonesix$, $\nthree$, $\nsix$, and $\none$).  The procedure
by which to average over $k$ is the same as described in
\S~\ref{res_effects_z}. The displayed error bars are the propagation of
the errors from the temporal statistics.   At these large values of $k$,
$\nueff$, $\etaeff$, $\pmeff$, and $\dkin/\dmag$ are relatively flat,
varying by a factor of at most 2.

The numerical viscosity and resistivity decrease as a function of
resolution. The dashed lines in the two upper panels of the figure
show the line $\nueff, \etaeff \propto N_x^{-2}$.  The viscosity and
resistivity decrease slower than this with increasing $N_x$; we measured
$\nueff, \etaeff \propto N_x^{-1.3}$. The figure shows that the effective
Prandtl number is constant with resolution to within the error bars. There
appears to be a slight increase in $\dkin/\dmag$ with resolution, but
this trend is not definitive given the large uncertainties on the data.

One might expect $\nueff$ and $\etaeff$ to decrease with increasing
resolution since these terms arise from truncation error.  Linear wave
advection test problems with $\at$ have shown that the truncation error
converges at second order \cite[e.g.,][]{stone08}.  On this basis, one
would expect $\nueff, \etaeff \propto N_x^{-2}$.  We find a shallower
decrease with $N_x$, but MRI turbulence is a fully nonlinear system,
and one should not necessarily expect the same convergence behavior as
in a linear system.

\section{Discussion and Conclusions}
\label{conclusions}

We have carried out a series of local, unstratified shearing box
simulations with the recently developed $\at$ code to study the characteristics
of MRI driven turbulence.  $\at$ uses a second-order,
conservative, compressive MHD algorithm, which is significantly different
from the algorithms employed in many of the previous MRI studies.
In our work, we have run several standard models for comparison with
previous work, and characterized the numerical dissipation of the $\at$
code for the shearing box problem.  Furthermore, we have exploited the energy
conservation property of $\at$ to carry out a study of energy flow within
MRI-driven turbulence.

To compare with previous numerical results, we have investigated the
effects of different initial field geometries (uniform or sinusoidal
$B_z$), varying domain aspect ratio, and numerical resolution.  In all
of our simulations, the MRI is initiated and sustained over many orbits.
The time- and volume-averaged properties of the resulting turbulent flow,
such as stress levels and magnetic and kinetic energies, are consistent
with previous results.  As in previous work, we find that boxes containing
net vertical field saturate at higher amplitudes compared to those
without net fields.  The total stress is proportional to the magnetic
pressure with a constant of proportionality $\sim 0.5$, but is independent of the
gas pressure.  In the net field simulation, the gas pressure increases by
a factor of 100, due to thermalization of the turbulence, without affecting
the stress.  The consistency of these results with past work indicate that
these properties do not result from details of the employed algorithm.

Fourier analysis of the turbulence shows that the largest scales in the
box dominate the energetics.  In the presence of a net field, the amplitude 
of the spatial power spectra is largely independent of
resolution on the largest scales.  This is not true for the zero net
flux simulations however.  For those simulations, the amplitude decreases as resolution increases,
which is consistent with the overall resolution behavior.
For net field simulations, the averaged turbulent magnetic and kinetic
energies increase slightly with resolution, whereas for the zero net field
simulations, the energies decrease with increasing resolution roughly in
proportion to the grid zone size.  This apparent lack of convergence for
the zero net field shearing box simulations was previously demonstrated
by \cite{from07a} using the ZEUS code.

The net field simulation shows intermittent channel flows which cause
temporary increases in stress through amplification of large-scale
MRI modes.   The parasitic modes described by \cite{good94} destroy the
channel flow within about one orbit of time, but the rapid increase in
stress produces a subsequent increase in thermal energy.  The presence
of these discrete channel flow events is a consequence of the box
size---larger boxes do not experience
them---but we use their presence to study the subsequent energy flow
following a rapid increase in stress.

Because $\at$ evolves the total energy equation, magnetic and kinetic energy
losses due to numerical grid-scale effects are added to the internal
energy.  This makes $\at$ well suited to examining the turbulent energy flow and
subsequent dissipation.  The recurring channel flows
in the net flux model provide a sudden injection of energy into the
box by increasing the stress operating on the shearing boundaries of
the box.  The injected energy appears as heat after $\sim 0.2$ orbits.
This corresponds to a timescale $\Omega^{-1}$, which equals $L_z/c_{\rm s}$
where $c_{\rm s}$ is the initial soundspeed.  This timescale determines the amplitude
of the $\alf$ speed, $v_{\rm A}$, and its fundamental MRI wavelength, $\lambda_{\rm MRI}$; $L_z/c_{\rm s} \sim 
\lambda_{\rm MRI}/v_{\rm A}$. The timescale is thus on the order of the eddy turnover time, indicating that
dissipational heating is a local process and that energy is not carried
over large distances before it is thermalized.

In the fiducial zero net magnetic flux simulation, $\zone$, there are no
recurring channel modes, making it more difficult to trace the flow of
injected energy.  The analysis is further complicated by the presence
of compressive waves that dominate the time derivative of the thermal
energy, $\td$.  These waves are also present in the net field simulations,
but their amplitude is smaller relative to the larger turbulent kinetic
energy found with a net field.  A detailed examination of the components
of the internal energy equation indicate that the compressive waves do not
appear to contribute significantly to irreversible heating.  By averaging
$\td$ for the zero net flux simulation, we find a correlation of $\td$
with $\ein$ on the same timescale of $\sim$ 0.2~orbits.

In the net field simulation, the dissipation of magnetic energy is larger
than that for the kinetic energy, not unexpected as the ratio of the average
magnetic to perturbed kinetic energy is $\sim 3.4$.  But the ratio of the magnetic
to kinetic dissipation rate is roughly constant at $\sim 1.7$. The fact that the
ratio of dissipation rates does not equal the ratio of energies
may result from a couple of possibilities.  First, there could be a
net transfer of magnetic to {\it perturbed} kinetic energy as was
suggested in \cite{bran95}.\footnote{\S~\ref{sustained} shows that
there is in fact a net transfer of kinetic to magnetic energy.  However,
this kinetic energy includes the shear flow, and thus, this result
tells us nothing of the energy transfer between magnetic and {\it perturbed} kinetic energy.}
Second, the difference in the ratios could arise from the effective Prandtl number
being larger than one.  In particular, if $\qk \propto \nueff \delta v^2/2$ and
$\qm \propto \etaeff B^2/2$, then $(B^2/\delta v^2)(\qk/\qm) \sim \pmeff$. With the above
values for the energy and dissipation ratios, we find $(B^2/\delta v^2)(\qk/\qm) \sim 2$,
which is consistent with the determination of $\pmeff$ from the Fourier analysis (see
discussion below).  The agreement
between the two separate calculations of $\pmeff$ may be coincidental, but it is suggestive
of $\qk \propto \nueff \delta v^2/2$ and $\qm \propto \etaeff B^2/2$.

\begin{figure}
\plotfiddle{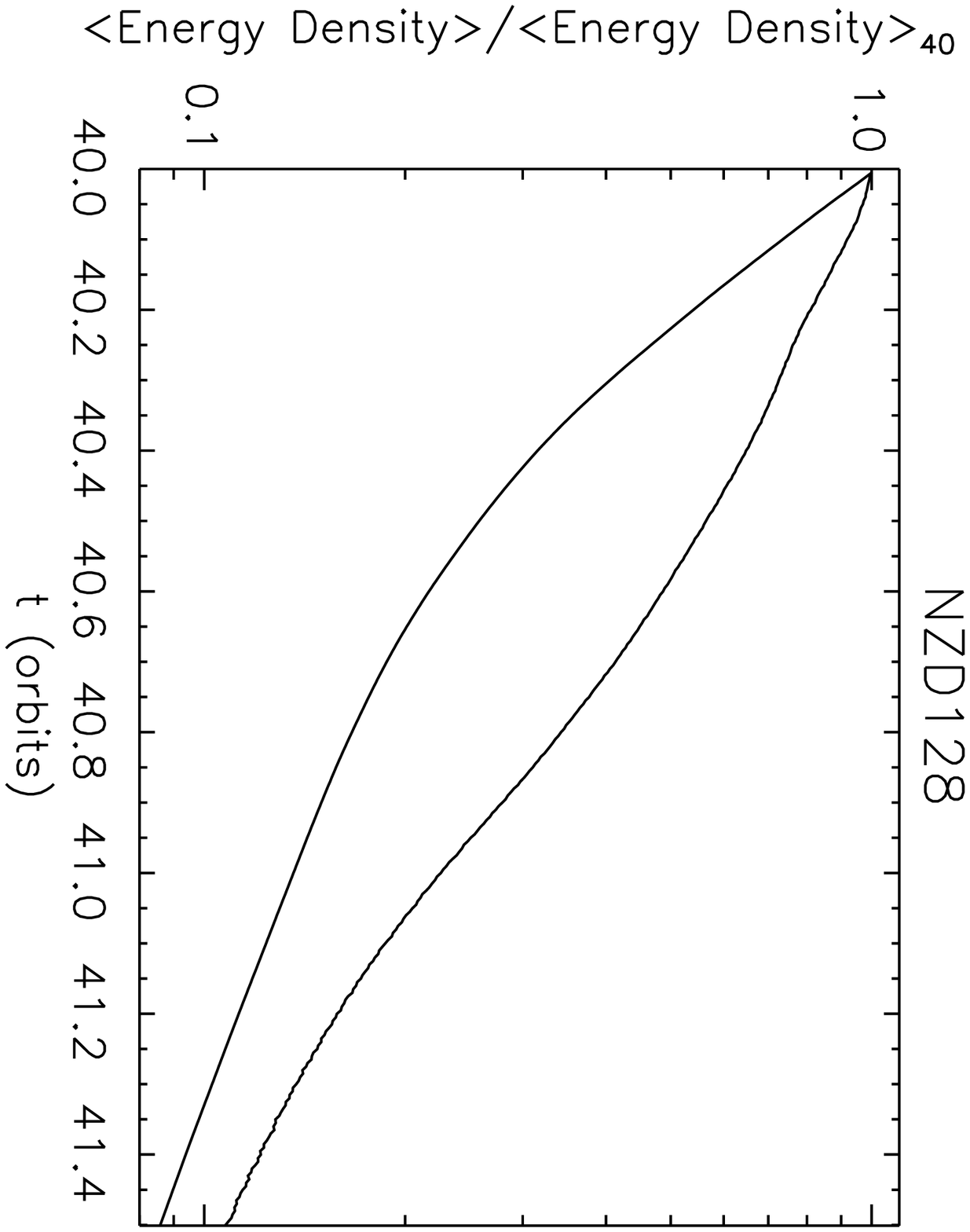}{1in}{90.}{200}{250}{-30}{-100}
\vspace{-3.12in}
\plotfiddle{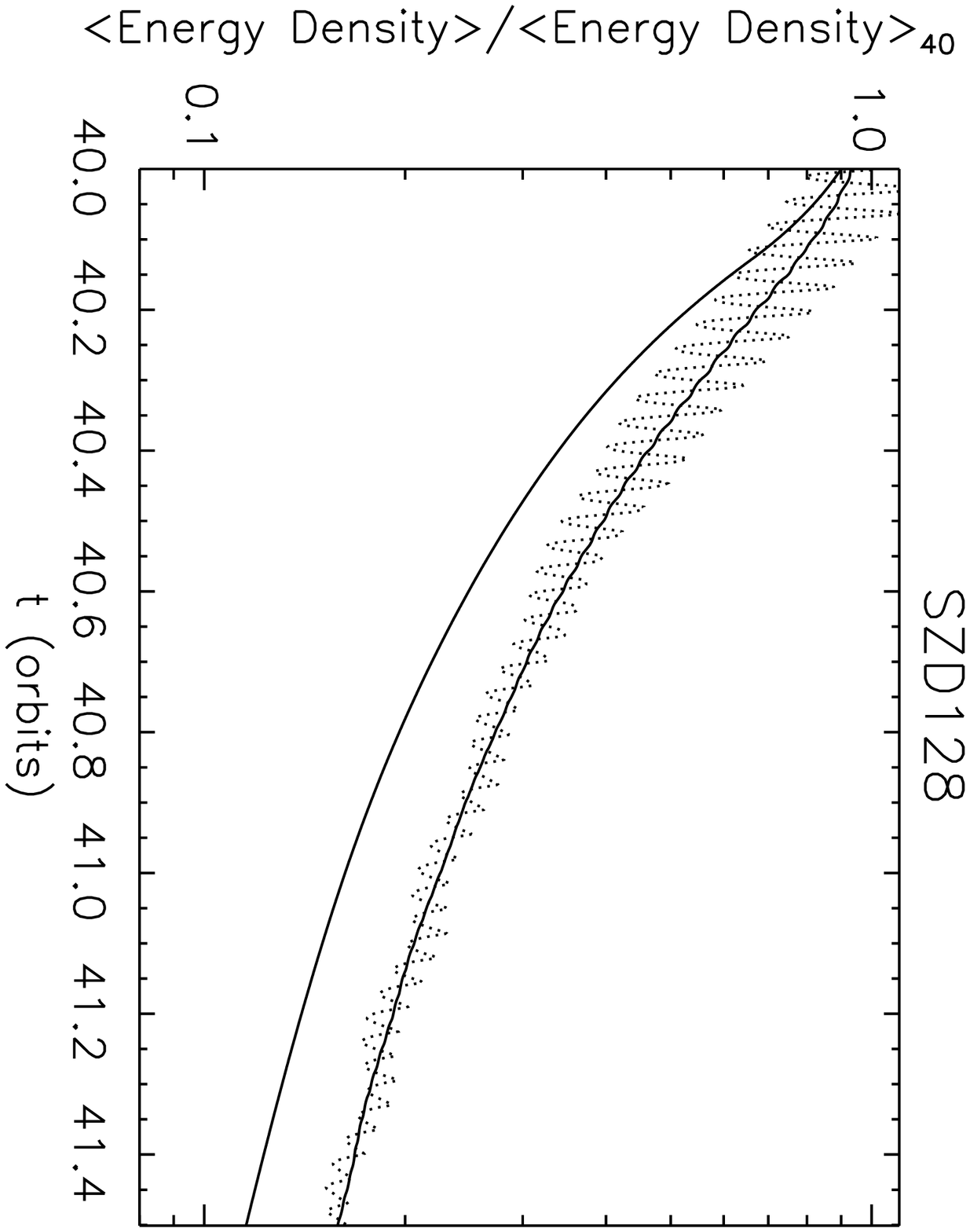}{0in}{90.}{200}{250}{235}{-100}
\caption{Volume-averaged magnetic and kinetic energy densities in the first 1.5~orbits of $\ndone$ (left) and $\zdone$ (right).  In both plots, the upper curves correspond to the kinetic energy density and the lower curves correspond to the magnetic energy density.  In $\zdone$, high frequency oscillations appear in the kinetic energy evolution.  To smooth away these oscillations, a moving window average was applied to the kinetic energy density.  The unsmoothed kinetic energy is shown by the dotted line, while the
smoothed kinetic energy is the solid line.  The magnetic energy density in the $\zdone$ plot has also been smoothed for consistency.
Both the kinetic and magnetic energy densities have been normalized to their respective (unsmoothed) values at $t =$~40~orbits.
\label{turb_decay}}
\end{figure}

The turbulence is sustained by the continued action of the MRI
in extracting energy from the differential rotation.  This can
be removed from the simulations allowing us to study the decay
of the turbulence in detail (simulations $\ndone$ and $\zdone$).
Figure~\ref{turb_decay} shows that magnetic
losses dominate over kinetic losses during this decay.  In both
simulations nearly 50\% of the magnetic energy and 20\% of the kinetic
energy has been dissipated after 0.2~orbits.  By one orbit into the decay,
most of the magnetic and kinetic energy has been lost.  Although these
decay timescales arise in a turbulent flow that lacks power input from
the MRI, the results are consistent with the conclusion that turbulent
energy dissipation occurs on a rapid timescale of order $\Omega^{-1}$.

\cite{from07a} used a detailed Fourier analysis
(\S~\ref{trans_funcs}) to study magnetic energy flow and thermalization
as a function of length scale in the shearing box.  In this analysis, the
individual terms in the evolution equation for the magnetic energy are
examined in Fourier space.  Averaging over time and assuming that the
magnetic energy is in a statistical steady state, one sets the sum of
these terms equal to a remainder, which is credited to numerical effects.
These numerical losses can then be modeled as an effective resistivity
(and viscosity for the kinetic energy), allowing one to characterize
the numerical dissipation in the simulation.

We repeated their analysis with $\at$ and extended it to the kinetic
energy.  The dominant effect at large scales is the generation of
magnetic field by the background shear.  This energy is transferred to
other scales by the turbulence.  Net positive field creation
by the turbulent flow and energy gains by the transfer between scales only
happens at small wavelengths.  This point of transition from loss to gain
happens at smaller scales for the zero net field simulation compared to the
net field model.  Magnetic dissipation dominates over kinetic dissipation
at small scales (i.e., $k L/(2\pi) \gtrsim 20$).  Modeling these as
an effective resistivity $\eta$ and viscosity $\nu$ shows that $\eta$
and $\nu$ drop with increasing resolution with a power that lies between
first- and second-order in grid resolution.  The effective Prandtl number,
on the other hand, is nearly constant as a function of resolution with
a value between $\sim 1.5$ and 2.

\cite{from07a} observed what they described as ``negative" resistivity
in an analysis restricted to the poloidal field alone.  In repeating their
exact analysis with $\at$, we also observed such an ``anti-dissipation"
at large scales.  This indicates that this effect is not associated with
a numerical algorithm limitation associated with ZEUS.  More likely,
it arises from the statistical uncertainty at large scales and from
the failure of the assumptions that go into the definition of the
dissipation term.  We note that the inclusion of the toroidal field $B_y$
in the analysis shows net dissipation at all scales, although again the
statistical variation is large at large scales.

In conclusion, what do these results imply for shearing box simulations
and the MRI?  First, as observed by  \cite{from07a}, the scales over
which turbulent energy generation occurs are not well-separated from
those where there is significant dissipation; the MRI operates over a
wide range of scales.  The MRI grows at a rate $\sim k v_{\rm A}$ for all $k$
less than $\Omega/v_{\rm A}$.  At large scales, a weak field will grow more
slowly than the timescale over which energy is transferred between
scales, between magnetic and kinetic forms, and ultimately thermalized.  If a field is
chopped up by reconnection, it may be reduced to small scales where the
MRI no longer operates.  In the presence of a net field, there will always
be a significant driving term at the scales set by that imposed field.
In the absence of such a field, however, the outcome will be determined
by the complex interplay of loss due to dissipation and amplification by
the MRI.  In the numerical simulations with zero net field, increasing
the resolution causes an overall decrease in the saturation energies.
\cite{from07a} attribute this to higher resolution enabling the MRI
to operate at intermediate scales which facilitates the transfer
of energy to small scales and promotes reconnection and dissipation.
What is perhaps surprising is that resolving the MRI at these scales leads
to greater field dissipation than would otherwise be accomplished by the
numerical losses that would occur if those scales were underresolved.
Because the same effect is observed with both $\at$ and ZEUS, it seems
likely that this ability of the MRI to transfer energy away from the
largest scales in the shearing box and to increase the total dissipation
is a physical rather than numerical effect. 

In related work, \cite{from07b} and \cite{lesur07} studied the effect of
varying the physical (not numerical) magnetic Prandtl number, $P_m$, on
the turbulence.  They found that the saturation amplitudes were increased
with increased $P_m$.  \cite{from07b} found evidence that there exists
a critical $P_m > 1$ below which zero net field simulations would die
out rather than achieve a steady turbulent state.  Our results in this
investigation show that this Prandtl number dependence is a distinct
effect from the observed dependence of the turbulence on resolution.
We find the numerical $P_m$ to be largely independent of resolution
in $\at$.  Taken together, however, the dependence on physical $P_m$
and the dependence on resolution point to the importance of small and
intermediate scale magnetic dissipation and reconnection to establishing
saturation amplitudes in MRI-driven turbulence.

As discussed by \cite{from07a}, numerical dissipation can deviate
significantly from physical dissipation.  In \S~\ref{fid_z}, we showed
that $\etaeff$ and $\nueff$ are relatively flat at small scales,
suggesting a resemblance to physical dissipation.   However, consider the
numerical Reynolds number as calculated from equation~(\ref{re}) for our
zero net flux simulations.  For $N_x$ = 128, we found $\reeff \sim$ 12000,
and $\pmeff \sim$ 1.6 for all of zero net flux simulations.  From the
parameter space studies of \cite{from07b}, these values for the Reynolds
and Prandtl numbers correspond to marginal MRI turublence; that is,
they lie very close to the critical line between sustained and decaying
turbulence.  For $N_x$ = 64, $\reeff \sim$ 4100, and the Reynolds number
is even smaller for the lower resolutions.  These values are well within
the decaying turbulence regime, but we find active MRI turbulence in all
of our simulations. These results show that the effective Reynolds and
Prandtl numbers of $\at$ as measured at large wavenumbers does not apply
at smaller $k$ values where there are many grid zones per wavelength.
Thus, the Reynolds numbers and Prandtl numbers that we calculate should
be taken as a measure of the effective numerical dissipation of the code
and not equated to a flow with the same Reynolds and Prandtl number as
determined by a simple physical resistivity and viscosity.

This result highlights an uncertainty associated with any MRI simulation
that depends only on numerical rather than physical dissipation.
It is apparent that the numerical Prandtl number can play an important
role in determining the ratio of magnetic to kinetic dissipation.
More speculatively, the Prandtl number may also play a role in the
timescale over which thermalization occurs.  In the present study, we
found that both the thermalization timescale and the effective numerical
Prandtl number were largely independent of resolution.   However, the
turbulent energy thermalization timescales and properties we measure may
be subject to change when explicit dissipation is included. It will be
a very important next step in this work to include physical dissipation
and verify these results.

This work is only the first step in applying $\at$
to the problem of the energetics of MRI turbulence.  The present
study provides a calibration of the numerical dissipation, which will
be important in future studies that include explicit resistivity and
viscosity.  Furthermore, the unstratified shearing box has the virtue
of simplicity and allows a detailed study of MRI turbulence without
too many confounding factors, but it also may prove too limited for
predictive application to accretion flows.  The inclusion of vertical
stratification and radiative cooling are both straightforward extensions
to the present study.  The detailed diagnostics developed and applied
in this study should prove valuable in this planned work.

\acknowledgments

We thank Jim Stone, Steve Balbus, and Sebastien Fromang for useful discussions and suggestions
regarding this work.  We also thank the anonymous referee whose 
comments and suggestions improved this paper.  
This material is based upon work supported by NASA
Grants NNG04GK77G and NAG5-13288.  The simulations were run on the NCSA
TeraGrid IA-64 Linux Cluster (Mercury) and the University of Virginia
Astronomy Department Beowulf Cluster.

\clearpage

\clearpage

\end{document}